\documentclass[11pt]{article}
\pdfoutput=1
\usepackage{jheppub}

\usepackage{amsmath,mathbbold,bbm,bm,slashed,graphicx, hyperref}

\newcommand{\arXivid}[1]{\href{http://arxiv.org/abs/#1}{\tt arXiv:#1}}

\newcommand\jhep[3]{{\it J. High Energy Phys.\ }{\bf #1} (#2) #3}
\newcommand\jmp[3]{{\it J.\ Math.\ Phys.\ }{\bf #1} (#2) #3}
\newcommand\prd[3]{{\it Phys.\ Rev.\ }{\bf D #1} (#2) #3}
\newcommand\prl[3]{{\it Phys.\ Rev.\ Lett.\ }{\bf #1} (#2) #3}
\newcommand\pr[3]{{\it Phys.\ Rev.\ }{\bf #1} (#2) #3}
\newcommand\plb[3]{{\it Phys.\ Lett.\  B}{\bf #1} (#2) #3}
\newcommand\atmp[3]{{\it Adv.\ Theor.\ Math.\ Phys.\ }{\bf #1} (#2) #3}
\newcommand\nature[3]{{\it Nature}{\bf #1} (#2) #3}
\newcommand\forp[3]{{\it Fortschr.\ Phys.\ } {\bf #1} (#2) #3}
\newcommand\ijtp[3]{{\it Int.\  J.\ Theor.\ Phys.\  } {\bf #1} (#2) #3}
\newcommand\cmp[3]{{\it Comm.\ Math.\ Phys.\ } {\bf #1} (#2) #3}
\newcommand\npb[3]{{\it Nucl.\ Phys.\ } {\bf B #1} (#2) #3}

\def\sgn{\operatorname{sgn}}

\title{The Spin of Holographic Electrons at Nonzero Density and Temperature}

\author[a]{Christopher P. Herzog}
\author[b]{and Jie Ren}
\affiliation[a]{
YITP, Stony Brook University \\
Stony Brook, NY  11794, U.S.A.
}
\affiliation[b]{
Department of Physics, Princeton University\\
Princeton, NJ 08544, U.S.A.
}
\emailAdd{Christopher.Herzog(at)stonybrook.edu}
\emailAdd{jieren(at)princeton.edu}

\preprint{\begin{tabular}{r}
PUPT-2409 \\
YITP-SB-12-07
\end{tabular}}
\abstract{
We study the Green's function of a gauge invariant fermionic operator in a strongly coupled field theory at nonzero temperature and density using a dual gravity description. The gravity model contains a charged black hole in four dimensional anti-de Sitter space and probe charged fermions. In particular, we consider the effects of the spin of these probe fermions on the properties of the Green's function. There exists a spin-orbit coupling between the spin of an electron and the electric field of a Reissner-Nordstr\"om black hole. On the field theory side, this coupling leads to a Rashba like dispersion relation. We also study the effects of spin on the damping term in the dispersion relation by considering how the spin affects the placement of the fermionic quasinormal modes in the complex frequency plane in a WKB limit. An appendix contains some exact solutions of the Dirac equation in terms of Heun polynomials.
}


\begin{document}

\maketitle

\section{Introduction}

Through gauge/gravity duality \cite{mal97,gub98,wit98}, a charged spinor field in an asymptotically anti-de Sitter (AdS) spacetime in a classical limit can be used to model strongly interacting fermions in field theory.  While the start of this program can be traced back to refs.\ \cite{Henningson:1998cd,Mueck:1998iz} which solve the Dirac equation in pure AdS spacetime,  with refs.\ \cite{Lee:2008xf, liu09, cub09}
there has been a resurgence of interest in the subject  focused on fermions at nonzero charge density in the hope of modeling strongly interacting cousins of Fermi liquids.
These so-called non-Fermi liquids are believed to underly some of the interesting physics of heavy fermion compounds and high temperature superconductors.  Initial gauge/gravity duality studies focused on charged black hole backgrounds.  By tuning the parameters of the fermionic field, both Fermi liquid and non-Fermi liquid behavior can be obtained \cite{fau09}.

These holographic models of strongly interacting fermions appear to be delicate to construct.  The charged black hole is subject to a wide variety of potential instabilities in the zero temperature limit.  For example, if the field theory contains an operator dual to a charged scalar field in the bulk, the black hole can develop scalar hair in a holographic superfluid phase transition
\cite{Herzog:2008he,Gubser:2008px,Hartnoll:2008vx,har08}.  If a four fermion interaction term with the right sign is added to the Dirac Lagrangian, there can be a Bardeen-Cooper-Schrieffer phase transition at low temperatures in the bulk \cite{Hartman:2010fk}.  Even with no extra terms in the Lagrangian,
if the fermions have a large enough charge, the condensation of a Fermi sea in the bulk will modify the geometry, producing an electron star at low temperatures \cite{har10a,har11a,har11b}.  The field theories dual to these electron stars exhibit the usual Fermi liquid behavior.  In contrast, the non-Fermi liquid behavior found by \cite{fau09} occurs in the limit where the Fermi sea outside the black hole is very small.

The delicate nature of these systems aside, they seem to be promising toy models to address some of the questions surrounding strongly correlated electron systems.  The current paradigm surrounding these toy models appeals to ideas of confinement in large $N$ gauge theories \cite{Iqbal:2011in, sac11, Hartnoll:2011pp, sac11b}.  The charge of the black hole should be carried by deconfined degrees of freedom -- non-gauge invariant fermions behind the horizon.  The added Dirac field is dual to a confined degree of freedom, i.e.\ a gauge invariant fermion or mesino.  Just as in QCD where the mesons and baryons interact weakly with each other in a large $N$ limit, these mesinos form a Fermi liquid which by definition is effectively weakly interacting.  It is the conjectured fermions behind the horizon which lead to non-Fermi liquid like behavior.

Our goals in this paper are modest.  We would like to provide a more careful consideration of spin physics and spin-orbit coupling in these holographic systems than has appeared heretofore in the literature.
A qualitative discussion of spin-orbit effects appears in ref.\ \cite{Benini:2010pr} in the context of coupling fermions to a d-wave holographic superconductor, but we shall try to be more quantitative and precise here.  By spin orbit coupling, we mean that a bulk charged fermion moving perpendicular to an applied electric field experiences an effective magnetic field that splits the degeneracy between the two spin states.

\begin{figure}
\centering
a) \includegraphics[width=0.32\textwidth]{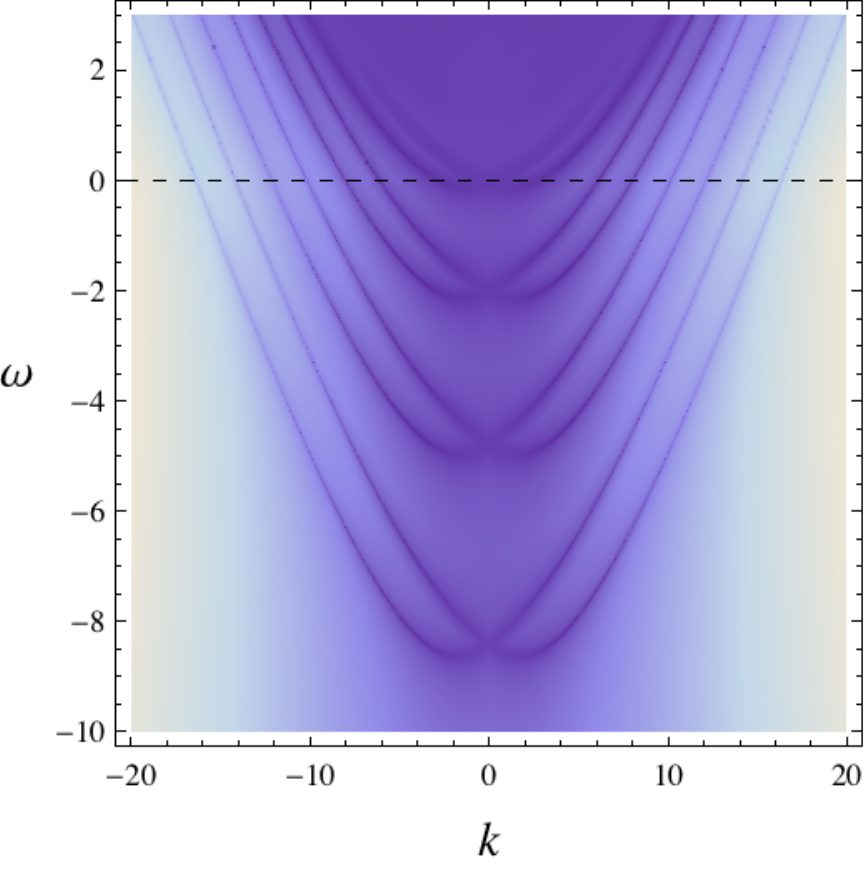}\quad
b) \includegraphics[width=0.32\textwidth]{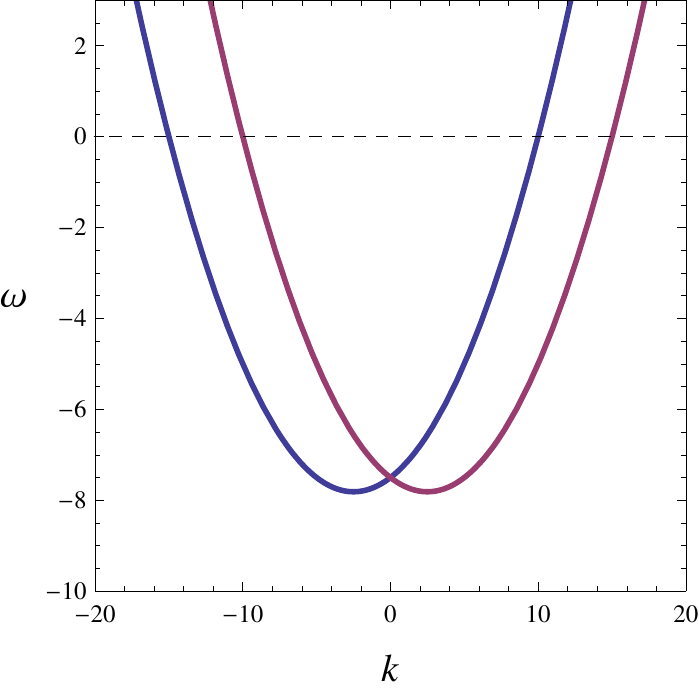}\quad
\caption{\label{fig:m2p}
(a) The dispersion relation in the boundary theory for a bulk fermion with charge times chemical potential divided by temperature $\mu q / 4 \pi T= 25$ and mass times AdS radius $m L=2$.
(b) The dispersion relation for a fermion with a Rashba type coupling.  The Fermi surface in both cases is indicated by the dashed line.
%The dispersion relations for two spins at different dipole coupling. The bottom plots are the combined bands with spin splitting. Increasing $\mu_p$ will cause bands to move toward upper left for the up spin, and toward upper right for the down spin. (As comparison, increasing $\mu_q$ will cause the bands to move downward.) The dashed line denotes the Fermi surface. Other parameters are $m=2$ and $\mu_q=25$.
}
\end{figure}

Let us start with an electrically charged black hole in $AdS_4$ at nonzero temperature to which we add a  spinor field.  Using the standard gauge/gravity duality dictionary, we may compute the quasinormal mode (QNM) spectrum of the spinor field which will allow us to deduce where the retarded Greens function for the dual gauge invariant fermionic operator has poles \cite{son02, iqb09}.
When the charge of the black hole is large enough, the imaginary parts of many of these quasinormal modes will be small, and we plot in figure \ref{fig:m2p}a
the real part of the freqency of these modes versus momentum.
One may think of these curves as dispersion relations for fermionic quasiparticles in the field theory.  Alternately, one may think of these curves as locations where the field theory may have a nonzero density of states.  (To compute the actual density of states, we need the full Green's function including the residues of the poles, and these residues may vanish at special points.)
The system is rotationally symmetric, and one can envision the full $k$ dependence by rotating the graph around the $\omega$-axis.
The details of this numerical computation are presented in section \ref{sec:example}.

The similarity between figure \ref{fig:m2p}a and figure \ref{fig:m2p}b is one of the central observations of this paper.  Figure \ref{fig:m2p}a resembles four copies of figure \ref{fig:m2p}b.  Figure \ref{fig:m2p}b is the dispersion relation for a nonrelativistic two dimensional electron gas with a spin orbit (or Rashba) coupling.  The Rashba Hamiltonian can be written
\begin{equation}
H = \frac{k^2}{2m_{\rm eff}} - \lambda \vec \sigma \cdot (\hat z \times \vec k) - \mu\,,
\label{rashba}
\end{equation}
where $\lambda$ is the Rashba coupling constant, $\mu$ a chemical potential, $\vec \sigma$ the Pauli matrices, $m_{\rm eff}$ the electron effective mass, and $\hat z$ the unit vector perpendicular to the gas.

From the bulk spacetime point of view, the similarity between these two figures is straightforwardly explained.  Electrons with the dispersion relation (\ref{rashba}) can be produced by a two dimensional slab-like geometry with a strong electric field perpendicular to the slab.  On the gravity side, the charge of the black hole provides the electric field.
The slab lies between the boundary of AdS on one side and the horizon on the other.  More precisely, deriving a
Schr\"odinger equation for the spinor field, one finds a potential barrier between the well in which the spinors live and the horizon.  Tunneling through the barrier produces the small negative imaginary part of the QNMs.

For the 2+1 dimensional field theory, this similarity naively presents a puzzle.
The usual derivation of the Hamiltonian (\ref{rashba}) is intrinsically 3+1 dimensional and relies on the presence of the electric field and a notion of spin-orbit coupling.
In the context of heavy fermion compounds and strange metals, one anticipates that spin should be essentially an internal $SU(2)$ symmetry of the electrons.  There are no strong magnetic or electric fields in these compounds, and the Fermi surface or surfaces should be spin degenerate.  There is no obvious mechanism for breaking the $SU(2)$ symmetry of these strongly interacting fermions at nonzero density.

The solution to this puzzle is that by the rules of the AdS/CFT correspondence,
the dual field theory is intrinsically relativistic.  Spin is not an internal symmetry but instead implies a corresponding transformation rule under the Lorentz group.  In 2+1 dimensions, angular momentum dualizes to a scalar.
Massive free fermions satisfying the Dirac equation $(\gamma^\mu p_\mu - im)\psi = 0$ carry a spin determined by the sign of their mass $\frac{1}{2} \sgn(m)$ (see for example \cite{Boyanovsky:1985qr}).\footnote{%
 In our conventions,  the gamma matrices obey $\{\gamma^\mu, \gamma^\nu \} = 2 \eta^{\mu\nu}$ where $\eta^{\mu\nu} = (-++)$.
 }
 In contrast,  massless fermions, because the little group is too small, carry no spin at all.
  We believe that the fermions in our field theory are massless for two reasons.  The first is that the field theory is conformal and a mass term breaks scale invariance.  The second is that while a mass term for the fermions breaks parity in 2+1 dimensions, the state we consider in the field theory appears to be parity invariant.  In support of this claim, note that the Hamiltonian (\ref{rashba}) is invariant under the parity operation that sends $(k_x, k_y) \to (-k_x, k_y)$ and $\psi \to \sigma_x \psi$.

There is an alternate intrinsically 2+1 dimensional way of motivating the Hamiltonian (\ref{rashba}).
Our black hole background is dual to a conformal theory at nonzero chemical potential $\mu$ and temperature.
The presence of energy and charge density identifies a preferred Lorentz frame $u^\mu = (1,0,0)$.
From an effective field theory point of view, it is natural to expect a modified Dirac equation of the form
\cite{Weldon:1982bn}
\begin{equation}
\left[ (1 + F) p_\mu + (-\mu + G)  u_\mu \right]  \gamma^\mu \psi = 0\,,
\label{effDirac}
\end{equation}
where $p^\mu = (\omega, k_x, k_y)$ is the four momentum and
$F$ and $G$ are arbitrary functions of $\omega$ and $k = \sqrt{k_x^2 + k_y^2}$.
Note we have not included a bare mass in this expression.
Choosing gamma matrices $\gamma^t = i \sigma_z$, $\gamma^x = \sigma_x$ and $\gamma^y = \sigma^y$ and
setting $F$ and $G$ to zero, we recover (\ref{rashba}) without the $k^2$ term and with $\lambda = 1$.
(In other words, the Rashba coupling itself is the Hamiltonian for a massless relativistic fermion in 2+1 dimensions.)
To add the $k^2$ term, we may posit that $G(\omega , k) \sim k^2$ which is allowed by the symmetries.
To get the different bands in figure \ref{fig:m2p}a, we may additionally posit the existence of several species of massless fermions with different charges $q_i$, replacing $\mu$ with $\mu q_i$ in (\ref{rashba}).  Ideally, we would like to derive this effective Dirac equation from an action, but we do not know how.

 One may object on technical grounds to the nonextremal black hole background used to produce figure \ref{fig:m2p}a.  From earlier work on the electron stars \cite{har10a,har11a,har11b}, for the large charge, low mass fermion chosen, it is clear that the bulk Fermi sea will cause a back reaction of the black hole background.  In principle, one should use the numerically computed electron star metrics to produce figure \ref{fig:m2p}a.   However, the phase transition between the charged black hole and the electron star is third order \cite{har11a}, and there are many qualitatively similar features between the charged  black hole
 and electron star backgrounds.  Using the numerical electron star metrics, which are only exact in an Oppenheimer-Volkoff approximation,
 should not change the story in a qualitative way.\footnote{%
 See \cite{sac11, Allais:2012ye} for progress going beyond the Oppenheimer-Volkoff approximation.
 Note that in the Oppenheimer-Volkoff approximation, the electrons are heavy and
 the spin splitting is very small.
 }

We begin this paper by revisiting the Dirac equation in these charged black hole and electron star backgrounds in section \ref{sec:dirac}.  We show, using the Pauli-Lubanski pseudovector, how to identify the different spin components of the fermion.  Next in section \ref{sec:QNMs}, we review how to compute the QNMs of a spinor field in these backgrounds, and connect the QNMs to poles of the retarded Green's function in the dual field theory.
Section \ref{sec:schrodinger} reviews how to convert the Dirac equation into an effective Schr\"odinger equation for the spinor for use in a WKB approximation.
The WKB limit will give us qualitative insight into the nature of this spin-orbit coupling.
Finally, in section \ref{sec:example}, we employ the machinery set up in the earlier sections to compute the QNMs of the spinor in a charged black hole in $AdS_4$, giving the details behind figure \ref{fig:m2p}a.
Section \ref{sec:numerics} contains a discussion of the numerical solution of the Dirac equation, while \ref{sec:wkb} discusses the Dirac equation in a WKB limit.
The WKB calculation contains some unusual features which we discuss at length.
We are able to show how the WKB approximation gets the sign of the imaginary part of the quasinormal modes correct.
%Appendix \ref{app:langer} justifies the Langer modified potentials we use in our WKB approximation.
Making use of Heun polynomials, appendix \ref{app:Heun} contains some exact analytic solutions of the Dirac equation for a charged spinor in a black hole in $AdS_5$.

\vskip 0.1in

\noindent
{\bf Note Added:}
After this paper was completed, we became aware of \cite{Alexandrov:2012xe} which has overlap with ours.

\section{The Dirac Equation Revisited}
\label{sec:dirac}

To have a gauge invariant fermionic operator $O_\Psi$ in a dual field theory,
we consider a spinor $\Psi$ in a curved spacetime with the action
\begin{equation}
S_\Psi=-i\int d^{d+1}x\sqrt{-g}\, \overline{\Psi}(\gamma^\mu D_\mu-m)\Psi\,,\label{eq:action}
\end{equation}
where $\Psi$ is a spinor of mass $m$ and charge $q$, and
$\overline{\Psi}=\Psi^\dagger\gamma^{\underline{t}}$.
Vielbein indices are underlined, and related to coordinate
indices by $\gamma^{\underline{a}}=e^{\underline{a}}_{\;\;\mu}\gamma^\mu$. The covariant derivative is
\begin{equation}
D_\mu=\partial_\mu+\frac{i}{4}\omega_{\mu,\underline{ab}}\gamma^{\underline{ab}}-iqA_\mu\,,
\end{equation}
where $\omega$ is the spin connection, $\gamma^{\underline{ab}}=\frac{1}{2}[\gamma^{\underline{a}},\gamma^{\underline{b}}]$, and $A_\mu$ is a gauge field.

With AdS/CFT applications in mind, we make the following simplifying assumptions on the metric and gauge field.  We assume a translationally and rotationally invariant metric of the form
\begin{equation}
ds^2 = g_{tt} dt^2 + g_{xx} (dx^2 + dy^2 + \ldots ) + g_{zz} dz^2\,,
\end{equation}
where the diagonal metric components $g_{\mu\mu}(z)$
depend only on a radial coordinate $z$.
 Additionally, we assume that $A_t$ is the only nonzero component of the vector potential and that it is a function only of the radial coordinate $z$.  In other words, there is a radial electric field whose strength may vary as a function of the radial direction.  The type of spacetimes we have in mind are Reissner-Nordstr\"om (RN) black holes and electron stars in AdS, both of which obey this set of assumptions.

Given these assumptions, the Dirac equation takes a particularly simple form.
The spin-connection term in the Dirac equation can be eliminated by using the
rescaled spinor $\tilde{\psi}=(-gg^{zz})^{1/4}\Psi$. The equation of motion for $\tilde{\psi}$ is
\begin{equation}
[\gamma^\mu(\partial_\mu-iqA_\mu)-m]\tilde{\psi}=0\,.
\label{eq:tilpsi}
\end{equation}
Translational symmetry in the directions orthogonal to $z$ suggest taking a Fourier transform
of $\tilde \psi$:
\begin{equation}
\tilde{\psi}(x^\mu,z)=\int\frac{d^4k}{(2\pi)^4}e^{ik_\mu x^\mu}\psi_k(z)\,,
\end{equation}
where $k^\mu=(\omega,{\bf k})$.
Because of rotational symmetry, we assume without loss of
generality that the spatial momentum is in the $x$ direction.
By plugging a single Fourier mode $\tilde{\psi}\sim e^{-i\omega t+ikx}\psi(z)$ into eq.~\eqref{eq:tilpsi}, we obtain the equation of motion for $\psi$:
\begin{equation}
\left[-i\sqrt{-g^{tt}}\gamma^{\underline{t}}(\omega+qA_t)+\sqrt{g^{zz}}\gamma^{\underline{z}}\partial_z
+i\sqrt{g^{xx}}\gamma^{\underline{x}}k-m\right]\psi=0\,.\label{eq:dirac}
\end{equation}

Specializing now to $d=3$,
we make the following choice of gamma matrices
\begin{equation}
\gamma^{\underline{t}}=\begin{pmatrix}
i\sigma_2 & 0\\
0 & i\sigma_2
\end{pmatrix},\qquad
\gamma^{\underline{z}}=\begin{pmatrix}
\sigma_3 & 0\\
0 & \sigma_3
\end{pmatrix},\qquad
\gamma^{\underline{x}}=\begin{pmatrix}
\sigma_1 & 0\\
0 & -\sigma_1
\end{pmatrix},\qquad
\gamma^{\underline{y}}=\begin{pmatrix}
0 & -i\sigma_1\\
i\sigma_1 & 0
\end{pmatrix}.
\label{eq:gamma}
\end{equation}
This choice is
consistent with the conventions in \cite{gul10} and allows for an easy generalization to a spacetime of arbitrary dimension.
Because we have set momentum in the $y$ direction to zero, the Dirac equation for the four component spinor
$\psi$
decouples into equations for two-component spinors $\psi = (\psi_+, \psi_-)^T$:
\begin{equation}
\left[\sqrt{-g^{tt}}\sigma_2(\omega+qA_t)+\sqrt{g^{zz}}\sigma_3\partial_z
\pm i \sqrt{g^{xx}}\sigma_1k-m\right]\psi_\pm =0\,.\label{eq:dirac2}
\end{equation}

We argue that $\psi_\pm$ correspond to fermions with opposite spin.
To see the spin direction, consider the Pauli-Lubanski pseudovector
\begin{equation}
W_{\underline{a}}=\frac{1}{2}\epsilon_{\underline{abcd}}J^{\underline{bc}}P^{\underline{d}}\,,
\end{equation}
where $J^{\underline{ab}}=\frac{i}{4}[\gamma^{\underline{a}},\gamma^{\underline{b}}]$ and $P^{\underline{a}}=-iD^{\underline{a}}$. We will show that $\psi = (\psi_+, 0)$ and $\psi = (0, \psi_-)$
are eigenstates of $W_{\underline{y}}$.
Acting on $e^{-i\omega t+ikx}\psi(z)$,
$P^{\underline{a}}=-ie^{\underline{a}}_{\;\;\mu}g^{\mu\nu}D_\nu$ is given by
\begin{equation}
P^{\underline{a}}
%=-ie^{\underline{a}}_{\;\;\mu}g^{\mu\nu}D_\nu \Psi
=(\sqrt{-g^{tt}}(\omega+qA_t),\,-i\sqrt{g^{zz}}\partial_z,\,\sqrt{g^{xx}}k,\,0)\,.
\end{equation}
Thus we obtain $ W_{\underline{y}}=$
\begin{scriptsize}
\begin{equation}
\frac{i}{2}\begin{pmatrix}
-i\sqrt{g^{zz}}\partial_z & \sqrt{g^{xx}}k-\sqrt{-g^{tt}}(\omega+qA_t) & 0 & 0\\
\sqrt{g^{xx}}k+\sqrt{-g^{tt}}(\omega+qA_t) & i\sqrt{g^{zz}}\partial_z & 0 & 0\\
0 & 0 & i\sqrt{g^{zz}}\partial_z & \sqrt{g^{xx}}k+\sqrt{-g^{tt}}(\omega+qA_t)\\
0 & 0 & \sqrt{g^{xx}}k-\sqrt{-g^{tt}}(\omega+qA_t) & -i\sqrt{g^{zz}}\partial_z
\end{pmatrix}.
\end{equation}
\end{scriptsize}
We find then that the Dirac equation (\ref{eq:dirac2}) can be written in terms of the $W_{\underline{y}}$ component of the Pauli-Lubanski pseudovector:
\begin{equation}
W_{\underline{y}}\begin{pmatrix}
\psi_+\\
\psi_-
\end{pmatrix}=\frac{1}{2}\,m\begin{pmatrix}
\mathbbold{1} & 0\\
0 & -\mathbbold{1}
\end{pmatrix}\begin{pmatrix}
\psi_+\\
\psi_-
\end{pmatrix},
\end{equation}
which means that the spin of $\psi_+$ is in the $y$-direction, and the spin of $\psi_-$ is in the opposite direction. Because we assumed that the momentum is in the $x$-direction, and the system has rotational invariance, the direction of spin is always perpendicular to the plane defined by the momentum and the radial direction $z$.
The Dirac equation decouples into these spin eigenstates.
%In short, the pseudovector demonstrates a decoupling of the Dirac equation into spin eigenstates perpendicular to the plane defined by the momentum and the radial direction.

This decoupling of spin eigenstates is in good agreement with flat space, nonrelativistic intuition \cite{Benini:2010pr}.
%
%In flat space given spin-orbit coupling,
%we expect the Dirac equation to split into spin components perpendicular to the momentum and electric field.
%Recall that in an atom, spin-orbit coupling comes from the interaction between the electric field of the nucleus and the spin of an orbital electron.
Starting with a massive fermion moving in the $x$-direction,
the fermion in its own rest frame will experience both an electric field in the radial $z$-direction and a
magnetic field in the $y$-direction.  This magnetic field will induce an energy splitting between
the $y$-spin up fermion and the $y$-spin down fermion.
The WKB limit we consider later will further strengthen this non-relativistic intuition.
There is also a possible coupling between the curvature  and the spin.
%We will see these spin effects in more detail when we look at the Dirac equation in a WKB limit.

\subsection{From quasinormal modes in the bulk to dispersion relations in the boundary}
\label{sec:QNMs}

In this section, we would like to take advantage of the well known AdS/CFT
relation between QNMs
of the bulk spacetime and poles in the Greens functions of the dual field theory \cite{son02}.
To that end, we begin by discussing the QNM
boundary conditions for the Dirac equation (\ref{eq:dirac2}).

We will assume that our metric in the limit $z \to 0$ is asymptotically of anti-de Sitter form:
\begin{equation}
g_{tt} \sim g_{xx} \sim g_{zz} \sim \frac{1}{z^2}\,.
\end{equation}
There are two linearly independent solutions of the two component
Dirac equation which approach the boundary as
\begin{equation}
\chi_1 \sim
 \left(
\begin{array}{c}
z^m \\
0
\end{array}
\right)
\; \; \;
\mbox{and}
\; \; \;
\chi_2 \sim
 \left(
\begin{array}{c}
0 \\
z^{-m}
\end{array}
\right).
\end{equation}
%[[ distinguish $\chi_\pm$ based on spin? ]]
We apply the ``Dirichlet'' boundary condition that the $\chi_2$ solution vanish.
Using these boundary conditions, the relation between the scaling dimension of the dual operator
$O_\Psi$ and the mass of the spinor is $\Delta = m+d/2$ \cite{Henningson:1998cd}.  The unitarity bound on the scaling dimension restricts us to $m \geq -1/2$.\footnote{%
 Using the ``Neumann'' boundary condition, we may consider spinors with dimension $\Delta = -m+d/2$ and $m \leq 1/2$.
}

As the ``Dirichlet'' boundary condition is Hermitian, the quasinormalness of the modes comes from the boundary condition applied in the interior of the geometry.  We assume that $g_{tt} \to 0$ at some $z_h>0$, and at this horizon, the phase velocity of the spinor wave function is in the positive $z$-direction.  For example, for a non-extremal blackhole, we may take the time and radial metric components to vanish as
$-g_{tt} \sim g^{zz} \sim 4 \pi T (z_h-z)$ where $T$ is the Hawking temperature.  In this case, our ingoing boundary condition is
\begin{equation}
\psi_\pm \sim (z_h-z)^{-i \omega / 4 \pi T}
 \left(
\begin{array}{c}
1 \\
1
\end{array}
\right).
\end{equation}
One may also consider more exotic situations, for example the Lifshitz geometry in the interior of the electron star \cite{har10a}.

Solving the Dirac equation with ``Dirichlet'' boundary conditions at $z=0$ and ingoing boundary conditions at the horizon is only possible for a discrete set of complex frequencies $\omega$ called QNMs.  As we tune $k$ in the Dirac equation, a given QNM will trace out a curve in the complex $\omega$ plane.  Provided the imaginary part of the QNM is small, this curve is essentially a dispersion relation for a quasi-stable particle.

To relate the QNMs to poles in the fermionic two-point function in the dual field theory, let us briefly recall how to compute these two-point functions.  The first step is to solve the Dirac equation with ingoing boundary conditions at the horizon and arbitrary boundary conditions at $z=0$, yielding a solution of the form $\psi_\pm = b_\pm \chi_1 + a_\pm \chi_2$ for each spin component
where $a_\pm$ is interpreted as proportional to a source in the dual field theory and $b_\pm$ as an expectation value.
The retarded Green's function is in reality a $2 \times 2$ matrix, but our choice of momentum has diagonalized it.
Through the theory of linear response, the retarded Green's function is $G_R^{\alpha \beta} = i \delta_{\alpha \beta} b_\alpha / a_\beta $ \cite{gul10}.  If $a_\alpha$ vanishes, $G_R$ will have a pole.

\subsection{A Schr\"odinger form for the Dirac equation}
\label{sec:schrodinger}

While at least three papers have considered the WKB limit of the Dirac equation in this AdS/CFT context  \cite{fau09, har11b,Iqbal:2011in}, these papers have largely ignored the effects of the spin of the electron.  Our focus here shall be on the spin.

The WKB approximation in this case assumes that the dimensionful parameters $m$, $\omega$, $q A_t$, and $k$ are large compared to the scale $\partial_z \ln \psi_\pm$ over which the wave function varies.
We may capture this limit by introducing a small parameter $\hbar$ multiplying $\partial_z$ in the Dirac equation,
\begin{equation}
\left[\hbar\sqrt{g^{zz}}\sigma_3\partial_z-m+\sqrt{-g^{tt}}\sigma_2(\omega+qA_t)
\pm i \sqrt{g^{xx}}\sigma_1k\right]\psi_\pm=0\,,\label{eq:diracw}
\end{equation}
and then expanding in $\hbar$.

Our first step is to convert the Dirac equation into Schr\"odinger form.  As noted in this AdS/CFT context by \cite{fau09}, there is no unique procedure.
Given an equation in Schr\"odinger form, one has the usual freedom to reparametrize the coordinate $y = f(z)$ and rescale the wave function $\phi \to Z \phi$ subject to the constraint $Z^2 f'$ is a constant.  However, there is an additional functional degree of freedom associated with converting a Dirac equation into Schr\"odinger form;
one can introduce the scalar wave function $\phi_\pm$ such that
\begin{eqnarray}
\psi_\pm = \left(
\begin{array}{c}
\alpha\,  \phi_\pm+ \hbar \beta\,   \partial_z \phi_\pm \\
\gamma \,  \phi_\pm + \hbar \delta\,   \partial_z \phi_\pm
\end{array}
\right).
\end{eqnarray}
The condition that $\phi_\pm$ satisfies a Schr\"odinger type equation $-\partial_z^2 \phi_\pm + V \phi_\pm = 0$ puts three constraints on the functions $\alpha$, $\beta$, $\gamma$, and $\delta$, but leaves a family of potentials $V(z)$ parametrized by an undetermined function \cite{Linnaeus:2010zz}.
%further introduce a new variable $y = f(z)$ and rescale $\phi_\pm \to g \phi_\pm$ which will keep the differential equation in Schr\"odinger form provided $g^2 \, \partial_z f$ is  a constant.
%In total, we find a family of Schr\"odinger equations parametrized by two undetermined functions.

One simple choice is to select
$\phi_\pm$ to be proportional to the first component of the two component spinor $\psi_\pm$:
\begin{equation}
\psi_\pm =
\left(
\begin{array}{c}
\sqrt{Z_{\pm k}}\,  \phi_{\pm} \\
i \frac{\sqrt{g_{zz}} m \sqrt{Z_{\pm k}} \, \phi_\pm  - \hbar \, \partial_z (\sqrt{Z_{\pm k}}\,  \phi_\pm) }{ Z_{\pm k}}
\end{array}
\right)
\end{equation}
%where the $\partial_z$ acts only on the logarithm,
where we have introduced a normalization factor
\begin{equation}
Z_{k} \equiv (g_{zz})^{1/2} \left[ \sqrt{-g^{tt}}(\omega + q A_t)  -  \sqrt{g^{xx}}k\right].
\label{Zkeq}
\end{equation}
Given these substitutions,
$\phi_\pm$ satisfies a Schr\"odinger equation of the form
\begin{equation}
-\partial_z^2 \phi_\pm + V_{\pm k, m} \phi_\pm = 0\,,
\label{Schreq}
\end{equation}
where the potential function is
\begin{eqnarray}
V_{k,m} (z) &=&
%\frac{1}{\hbar^2} g_{zz} \left(
%m^2 + g^{tt}(\omega + q A_t)^2  +  g^{xx} k^2 \right) \nonumber \\
%Z_k^4
 \frac{1}{\hbar^2} \left( g_{zz}
m^2 - Z_{-k} Z_k \right)
% \nonumber \\
%&&
 -\frac{m g_{zz}}{\hbar}
 \frac{\partial_z ( \sqrt{g^{zz}} Z_k)}{Z_k}
%m \sqrt{g_{zz}} \, \partial_z \ln \left[ \sqrt{g^{zz}} Z_k^2 \right]
%
+  \sqrt{Z_k} \partial_z^2 \frac{1}{\sqrt{Z_k}}
%- \partial_z^2 \ln Z_k  +\left( \partial_z \ln Z_k \right)^2
 \label{diracpot}
 \,.
\end{eqnarray}
Note that the spin dependence of the potential enters only at subleading order.  At leading order in $\hbar$, $V_{k,m}$ is independent of the sign of $k$.

At leading order in $\hbar$, the spinor potential $V_{k,m}$
is exactly what one obtains for a charged scalar particle in this curved spacetime.
If we start with the
scalar wave equation $(D_\mu D^\mu - m^2) \Phi = 0$
and let $\Phi = e^{-i \omega t + i k x} Z(z) \phi(z)$ where $Z = \sqrt{g_{zz}} (-g)^{-1/4}$, then we obtain
$ - \partial_z^2 \phi+ V_s \phi = 0$ where
\begin{eqnarray}
V_s(z) &=& \frac{1}{\hbar^2} \left(  g_{zz} m^2 - Z_{-k} Z_{k} \right)
%\nonumber \\
%&&
+ Z \partial_z^2 \frac{1}{Z} \,,
%- \partial_z^2 \ln Z + (\partial_z \ln Z)^2 \ ,
\label{scalarpot}
\end{eqnarray}
and we have introduced factors of $\hbar$ analogously to the spinor case.

There is an important difference between the scalar and the spinor case.  The normalization factor $Z_{k}$ may vanish at a point in spacetime where the energy of the particle is equal to the local chemical potential plus a $k$ dependent correction.  At such a point, the spinor potential $V_{k,m}$ will have singularities at subleading order in $\hbar$.
We will see in the next section how these subleading singularities mean that while scalar QNMs can lie in the upper half plane, the spinor QNMs will not.

Before getting into the details, we can say something
about the number of zeroes $Z_k$ possesses in the interval $0 < z < z_h$.  In general, we will associate the boundary value of $A_t$ with a chemical potential: $A_t(0) = \mu$; at the horizon we set $A_t(z_h) = 0$.  Given the boundary conditions described in section \ref{sec:QNMs} for the metric $g_{\mu\nu}$, close to the horizon we find that
$Z_k \sim \omega / 4 \pi T ( z_h - z)$.  At the boundary, we find instead that $Z_k(0) = \omega + q \mu - k$.
Thus, if $\omega + q \mu - k $ and $\omega$ are of the same sign, there will be an even number of zeroes between the boundary and the horizon.  If they have opposite sign, there will be an odd number.  In the cases we consider below, $q\mu > k - \omega$, and so the parity of the number of zeroes is determined by the sign of $\omega$, odd if $\omega <0$ and even if $\omega > 0$.

Before continuing to an example, we would like to point out another simple choice of Schr\"odinger equation.
We can let  $\phi_\pm$ be proportional to the second component of $\psi_\pm$ rather than the first.
To compute this alternate Schr\"odinger equation, we do not need to do the calculation again.
Note instead that eq.~(\ref{eq:diracw}) is
invariant under $k \to -k$, $m \to -m$, and $\psi_\pm \to \sigma_1 \psi_\pm$.
Thus the Schr\"odinger equation for the second component is given by replacing
$V_{k,m}$ with $V_{- k,-m}$.  Interestingly, this symmetry implies that
the potentials $V_{k,m}$ and $V_{-k,-m}$ have the same quasinormal mode spectrum (being careful to exchange the boundary conditions on the two components of $\psi_{\pm}$ as well).
The existence of this second isospectral potential allows for some cross checks when we perform a WKB analysis below.

\section{An Example: AdS-Reissner-Nordstr\"om Black Hole}
\label{sec:example}

To study a fermionic system at
nonzero temperature and density, we use the RN black hole as the background geometry, solve the Dirac equation coupled to a U(1) gauge field, and obtain the QNMs.
We will solve the Dirac equation both numerically and using WKB.  Our main interest is the location of the poles of the Green's function.
Most of our results are for the $m=2$ spinor in $AdS_4$.

We consider a charged black hole in $AdS_4$ in the Poincar\'e patch for which the metric has the form
\begin{equation}
ds^2=\frac{L^2}{z^2}\left(-f(z)dt^2+dx^2+dy^2+\frac{dz^2}{f(z)}\right).\label{eq:ansatz}
\end{equation}
The conformal boundary is at $z=0$, and the horizon is at $z=z_h$, where $f(z_h)=0$. We set $L=1$ and $z_h=1$,
which are allowed by two scaling symmetries \cite{har08}. The charged black hole solution and its probe limit are as follows:
\begin{align}
& \text{RN-$AdS_4$:} & &f=1-\Bigl(1+\frac{\mu^2}{4}\Bigr)z^3+\frac{\mu^2}{4}z^4 & &\text{with}\quad A_t=\mu(1-z)\,;\label{eq:RN}\\
& \text{probe limit:} & &f=1-z^3 & &\text{with}\quad A_t=\mu(1-z)\,.\label{eq:probe}
\end{align}
We work in the probe limit $\mu=0$ and $\mu q = \mu_q$ held fixed.  In this limit, the electric field does not backreact on the geometry. If we use the full solution to the RN black hole, the qualitative features will not change, as we will discuss at the end of section \ref{sec:numerics}.

\subsection{Numerics}
\label{sec:numerics}

To solve the Dirac equation (\ref{eq:dirac2}) numerically in this spacetime, we approximate the ingoing solution by a Taylor series near the horizon $z=1$, integrate numerically to a point near the boundary $z=0$, and fit the numerical solution to the boundary expansion
$\psi_\pm = b_\pm \chi_1 + a_\pm \chi_2$.  The numerical integration was performed using Mathematica's NDSolve routine \cite{Mathematica}.
As we are interested in QNMs and the corresponding Green's function singularities, we focus on the value of the source $a_\pm$.

\begin{figure}
\begin{minipage}[t]{\textwidth}
\centering
\includegraphics[height=0.365\textwidth]{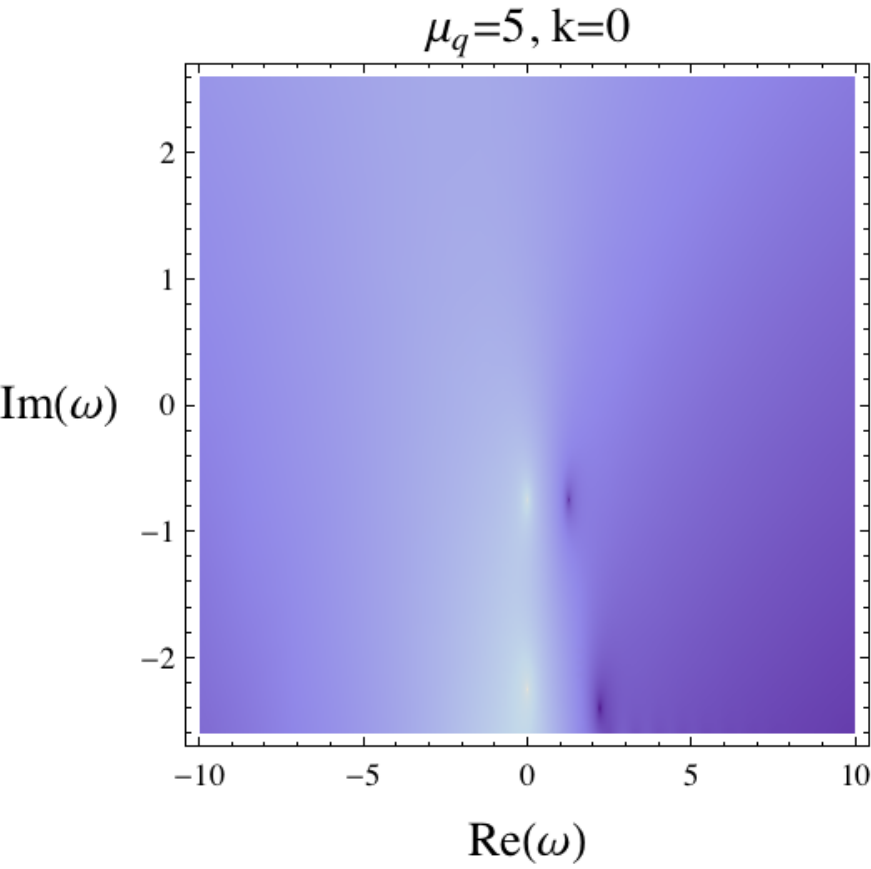}
\includegraphics[height=0.365\textwidth]{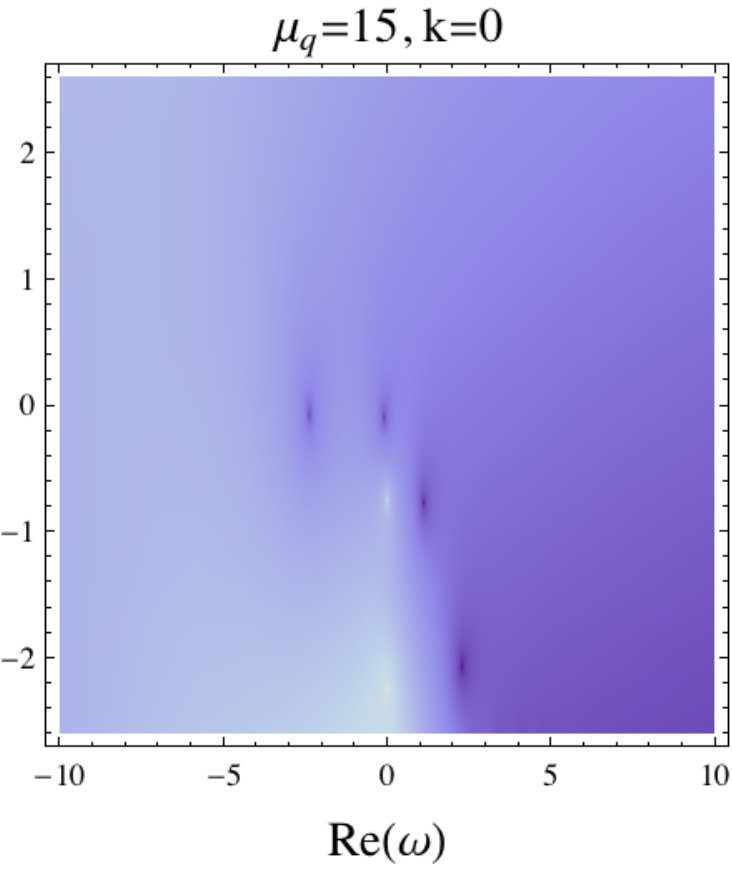}
\includegraphics[height=0.365\textwidth]{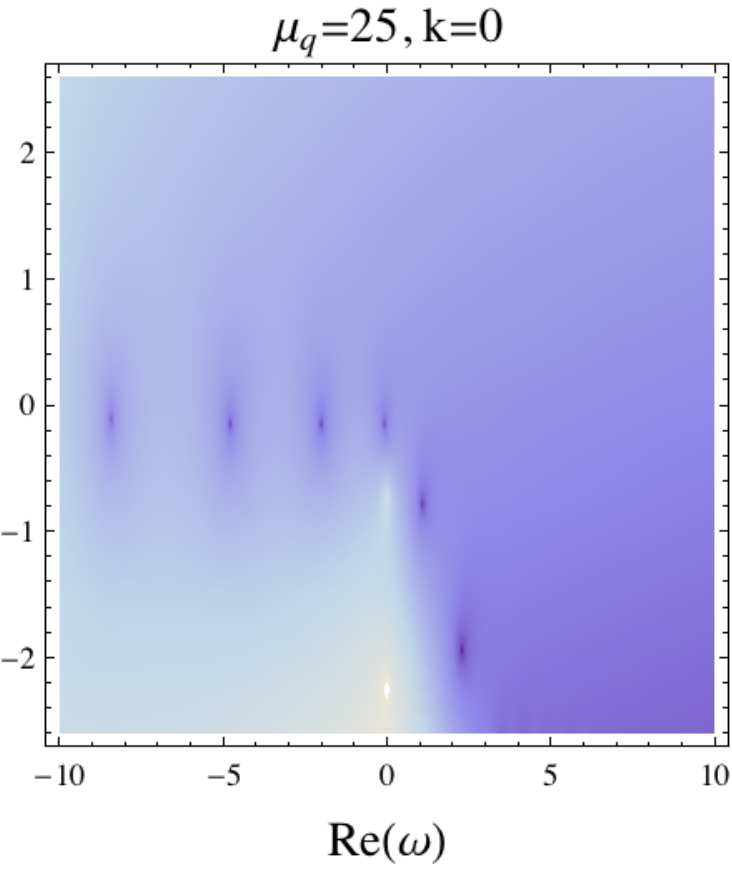}
\caption{\label{fig:m2q} Motion of the poles in the complex $\omega$ plane as we increase $\mu_q$. From left to right, $\mu_q=5$, $15$, $25$. Other parameters are $m=2$ and $k=0$.}
\end{minipage}\\[15pt]
\begin{minipage}[t]{\textwidth}
\centering
\includegraphics[height=0.365\textwidth]{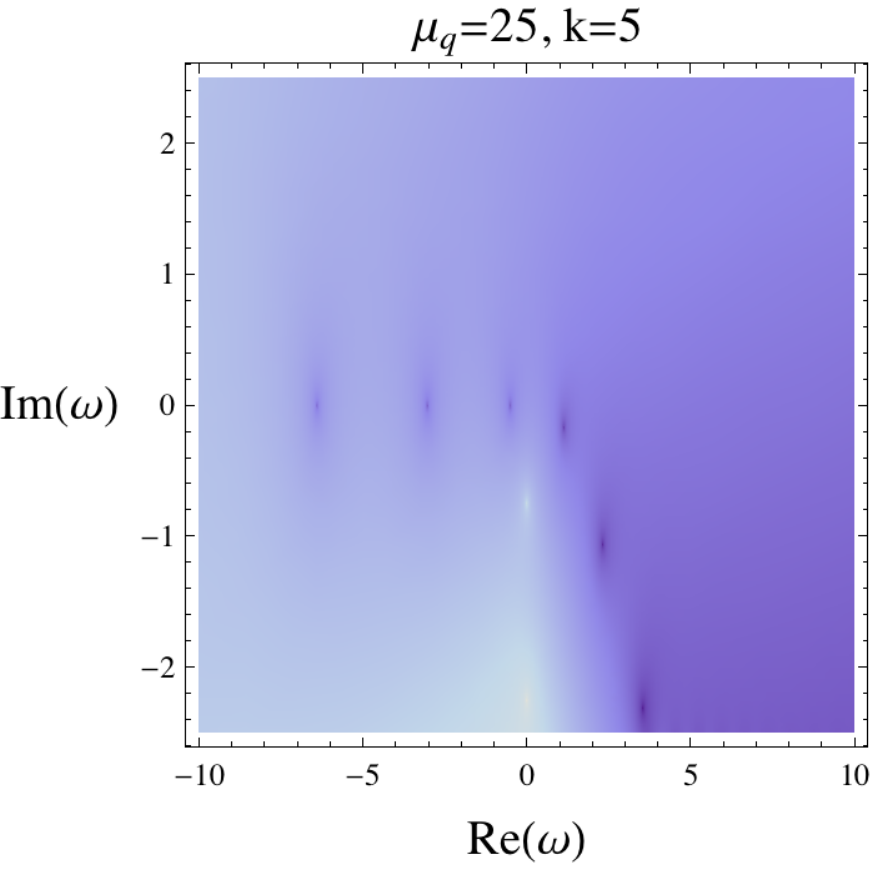}
\includegraphics[height=0.365\textwidth]{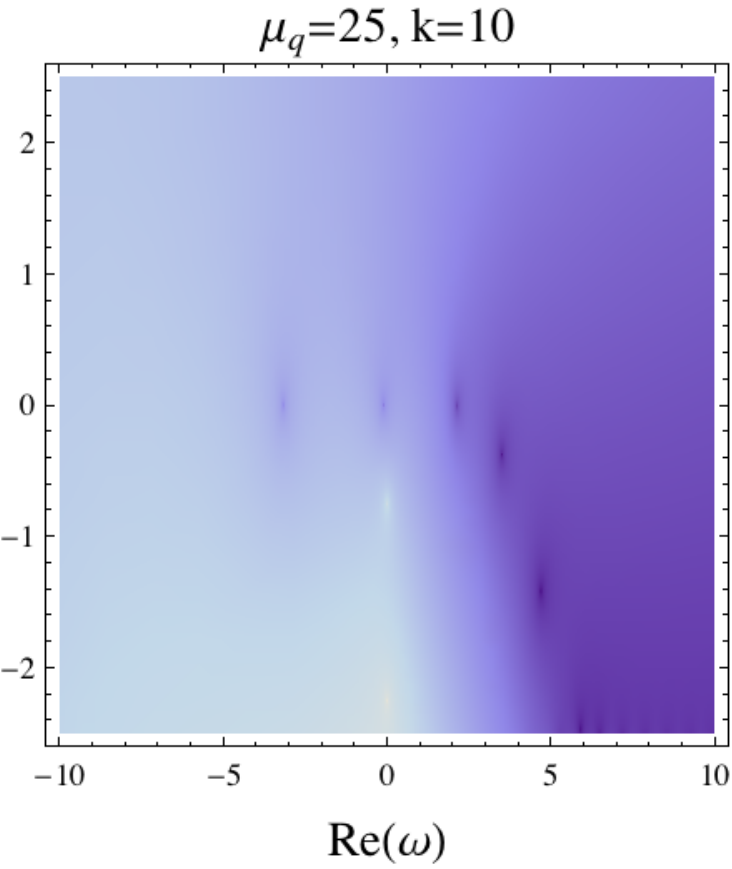}
\includegraphics[height=0.365\textwidth]{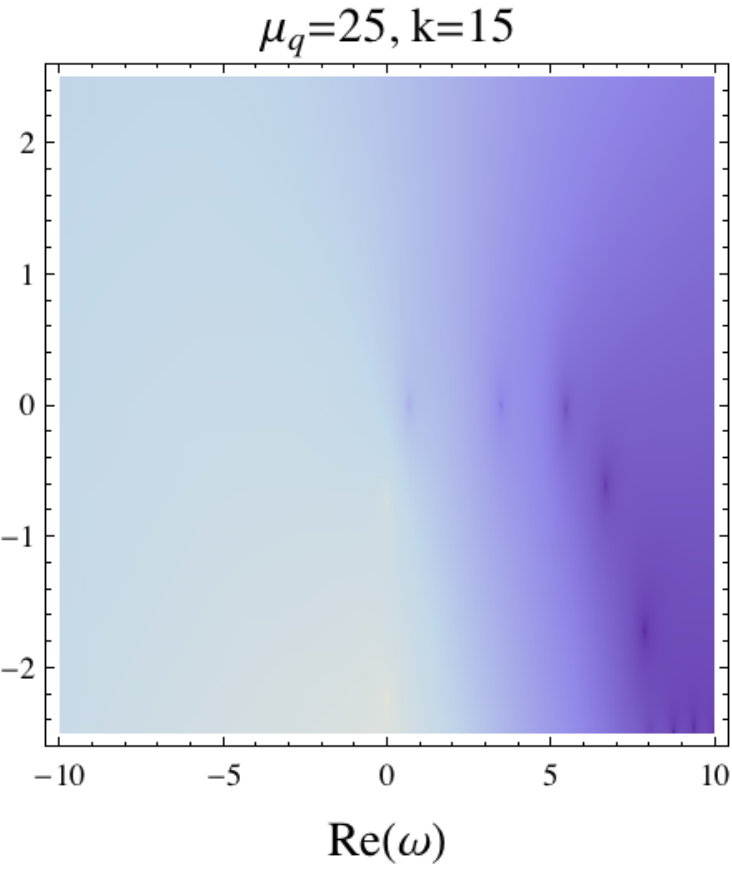}
\caption{\label{fig:m2k} Motion of the poles in the complex $\omega$ plane as we increase $k$. From left to right, $k=5$, $10$, $15$. Other parameters are $m=2$ and $\mu_q=25$.}
\end{minipage}
\end{figure}

Figure \ref{fig:m2q} presents a
 density plot of $|a_+|$ as a function of complex $\omega$. We
 see that the poles of the Green's function in the complex $\omega$
 plane have the following features:
\begin{itemize}
\item As we increase the chemical potential $\mu_q$, there are more and more poles along the negative
real $\omega$ axis.
Decreasing $\mu_q$ moves these poles to the right. Once they cross the imaginary axis, they begin to move away from the positive real axis.  See figure~\ref{fig:m2q}.
\item If we fix $\mu_q$ and increase the momentum $k$, the poles move to the right.
As $\omega=0$ corresponds to the Fermi surface,
when a pole crosses the imaginary axis, we obtain a Fermi momentum $k_F$. See figure~\ref{fig:m2k}.
\end{itemize}
We can obtain many Fermi momenta $k_F^{(n)}$ and many Fermi surfaces. The poles close to the real $\omega$ axis correspond to quasibound states, and the number of them at $k=0$ equals the number of Fermi surfaces. The number of Fermi surfaces grows linearly with the chemical potential, which can be seen by comparing figure~\ref{fig:m2q} and figure~\ref{fig:QNM50} in the next section (or equivalently tables (\ref{mu25k5}) and (\ref{mu50mu75})).

For the other spin ($k\to -k$), the poles will be in slightly different locations.
The spin splitting can be seen more obviously by looking at figure \ref{fig:m2p}.  The figure is a plot of the dispersion relation and was constructed by superposing density plots of $|a_+|$ and $|a_-|$ in the $k$-$\omega$ plane.
The density plot of $|a_+|$ corresponds to the parabolas on the left and $|a_-|$ to the parabolas on the right.

\begin{figure}
\centering
\includegraphics[width=0.35\textwidth]{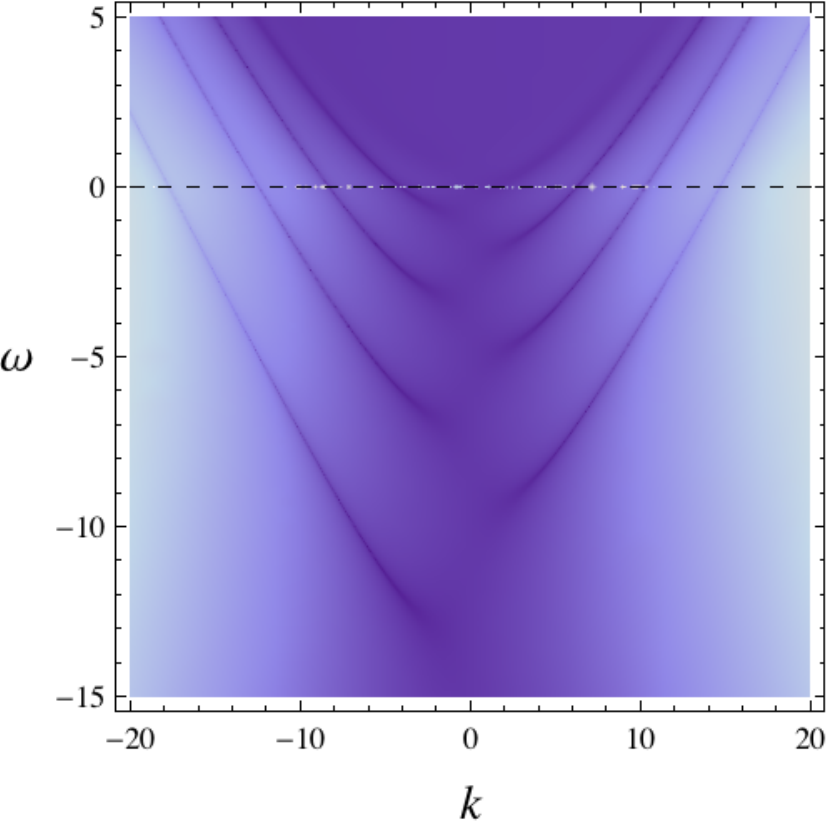}
\caption{\label{fig:m0} Dispersion relation for a massless fermion with $\mu_q=20$. Its mirror image $k\to -k$ is for the opposite helicity. There are no poles in the Green's function at $k=0$.}
\end{figure}

For comparison, we also plot the dispersion relation for a massless $m=0$ bulk fermion
in figure \ref{fig:m0}.  (Here only $|a_+|$ is plotted.
The density plot for the other spin component $|a_-|$ is the mirror image.)
Unlike the massive case, there do not seem to be well defined quasiparticles close to the $k=0$ axis.
For the $m=0$ case there are no poles in the Green's function at $k=0$ and $m=0$.
This fact can be checked analytically by solving the Dirac equation to obtain $G_R(\omega, k=0) = i$ \cite{fau09}.
From a bulk point of view, the absence of these poles is presumably related to the absence of a rest frame for a massless particle.
As can be seen in figure \ref{fig:m0}, poles do appear for $k \neq 0$.  For the field theory dual, the interpretation is more obscure.  Ref.\ \cite{Allais:2012ye} relates the absence to strong interactions with a background continuum of states existing inside an ``IR lightcone''.  Why then the interactions are suppressed for larger bulk masses still needs to be explored.

\begin{figure}
\begin{minipage}[b]{0.5\textwidth}
\centering
\includegraphics[width=0.7\textwidth]{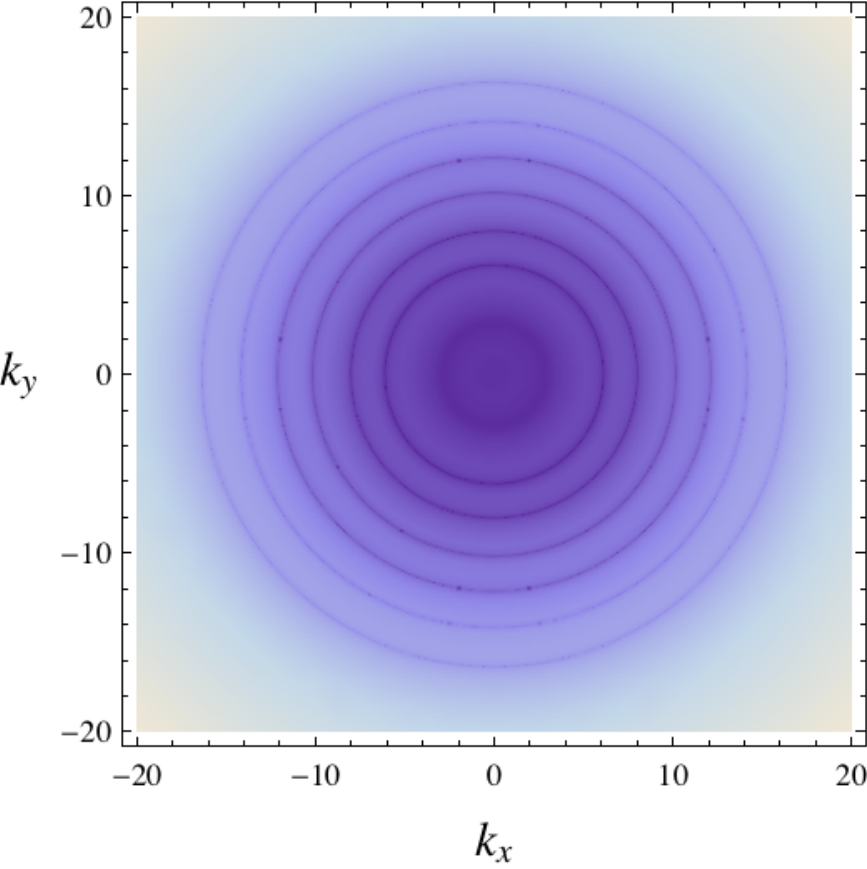}
\vspace{0pt}
\end{minipage}
\begin{minipage}[b]{0.5\textwidth}
\centering
\includegraphics[]{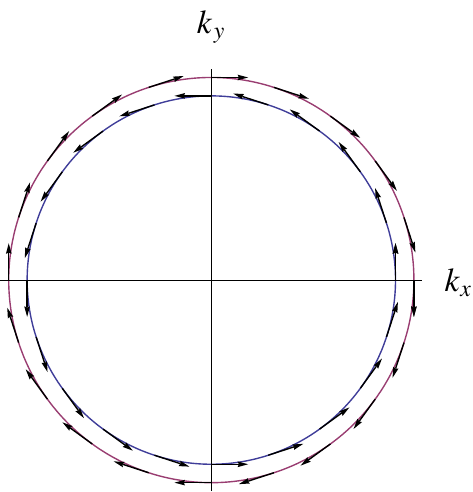}
\vspace{15pt}
\end{minipage}
\caption{\label{fig:FS} The left plot shows the Fermi surfaces with spin splitting. The parameters are $m=2$ and $\mu_q=25$. The right plot illustrates the directions of spin for massive bulk fermions.}
\end{figure}

Figure \ref{fig:FS} (left) shows the location of the Fermi surfaces in momentum space and also demonstrates the spin splitting effect.  As we discussed above, the spin of the bulk spacetime fermions is perpendicular to the momentum and the electric field.  Thus the bulk Fermi surfaces are spin polarized, as shown schematically in figure \ref{fig:FS} (right).
A similar effect has been studied in spin Hall systems \cite{sin03} and observed in experiments \cite{hsi10} for electrons that while confined to a plane still have a three dimensional spin.

Note that for the 2+1 dimensional field theory fermions, the spins shown in figure \ref{fig:FS} (right) are misleading.  From a purely 2+1 dimensional perspective, we argued in the introduction that the fermions are both massless and spinless.   The extra degrees of freedom producing the second Fermi surface come from the hole states.  The dispersion relation for the hole states has been bent upward, emptying out the infinite Fermi sea and producing a second Fermi surface.

\begin{figure}
\centering
\includegraphics[height=0.365\textwidth]{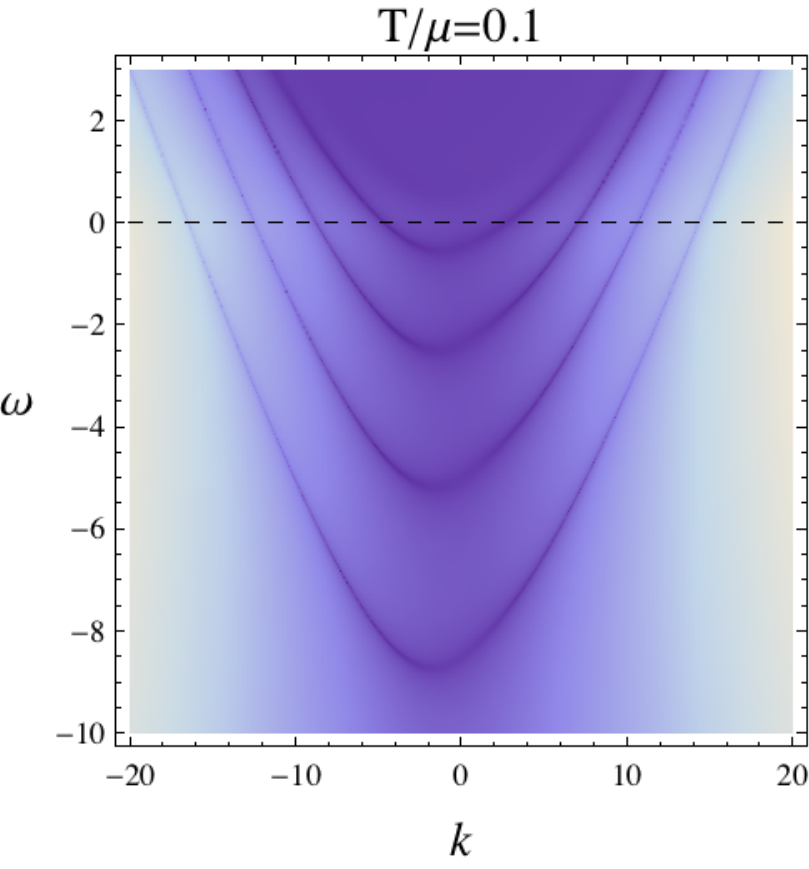}
\includegraphics[height=0.365\textwidth]{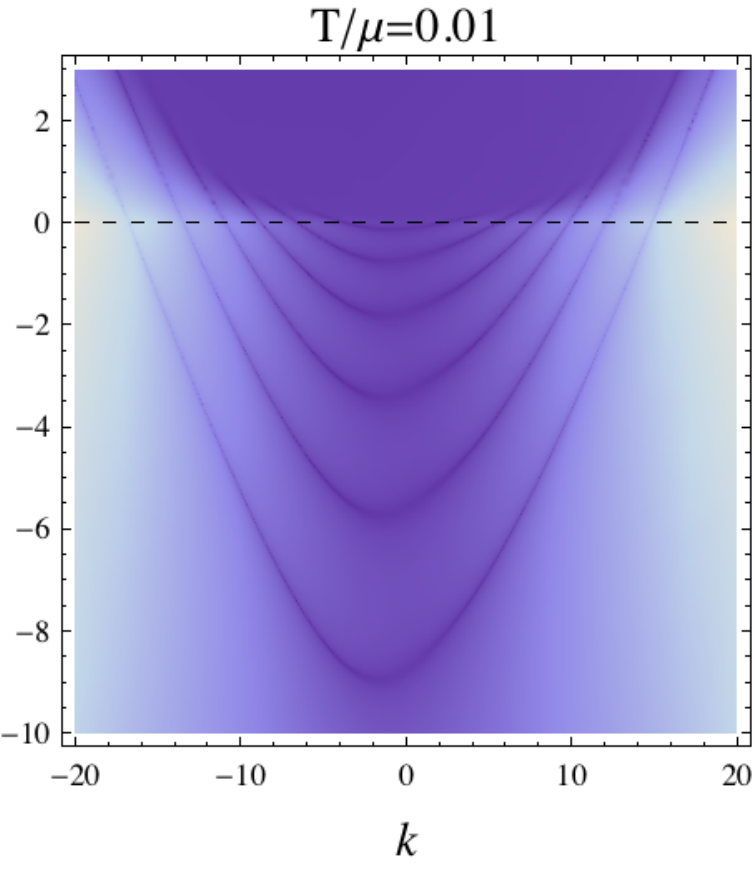}
\includegraphics[height=0.365\textwidth]{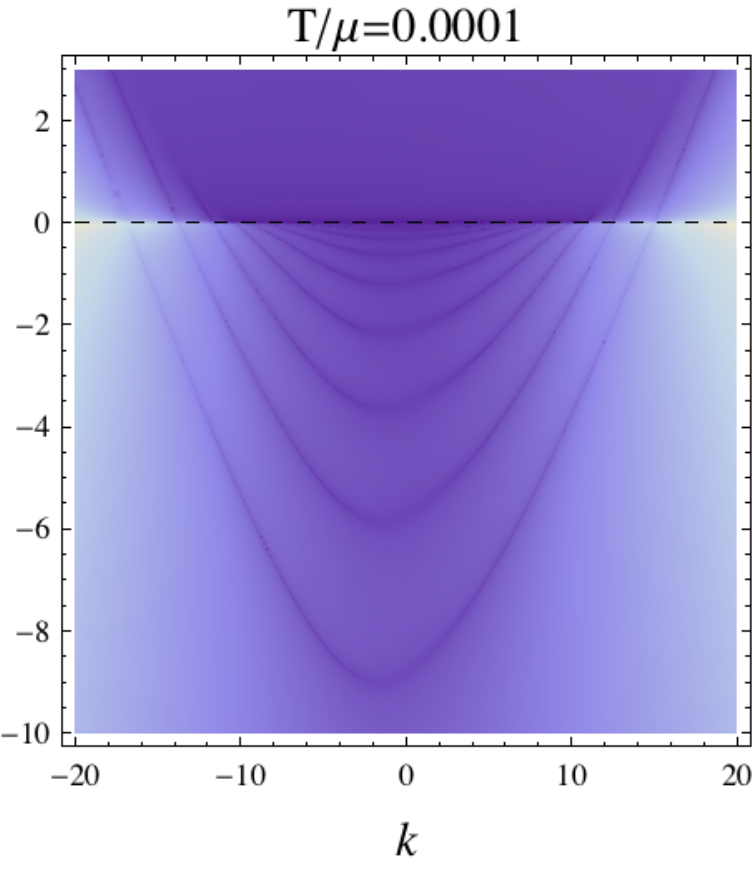}
\caption{\label{fig:m2T} Temperature dependence of the dispersion relation with fixed $\mu_q=25$. This is the only plot that uses the full solution of the RN black hole in this work. The chemical potential for the three plots are $\mu\approx 1.7665$, $3.2218$, $3.4615$ (from left to right). The extremal case corresponds to $\mu=2\sqrt{3}\approx 3.4641$.
These are spin up fermions.
}
\end{figure}

Before moving on to WKB, we promised a discussion of the validity of our probe approximation.
The temperature for the charged black hole solution eq.~\eqref{eq:RN} is
\begin{equation}
T=\frac{12-\mu^2}{16\pi}\,.
\end{equation}
If we lower the temperature by increasing the chemical potential $\mu$, more Fermi surfaces will appear, but the size of the outer Fermi surface will remain roughly the same, as figure~\ref{fig:m2T} shows. By comparing figure~\ref{fig:m2T} and figure~\ref{fig:m2p}a, we can see that if we consider the high temperature regime, i.e., $T/\mu>0.1$, there is no essential difference between the full RN solution and the probe limit in eq.~\eqref{eq:probe}.

%\begin{figure}
%\centering
%\includegraphics[width=0.5\textwidth]{poteff_m2q25.pdf}
%\caption{\label{fig:Veff} Effective potential for the Dirac equation with $m=2$, $k=0$, and $\mu_q=25$.}
%\end{figure}

\subsection{WKB}
\label{sec:wkb}

\begin{figure}
%%\begin{center}
%%\includegraphics[width=0.45\textwidth]{typicalthree.pdf} \quad
%%\includegraphics[width=0.45\textwidth]{typicalfive.pdf}
%%\end{center}
%%\caption{\label{fig:typicalthree}
%%The potential $V_{k,m}$ with three turning points (left) and five turning points (right).
%%The horizon is indicated by the dashed line.
%%}
%%\end{figure}
%\begin{minipage}[t]{\textwidth}
\centering
a) \includegraphics[width=0.35\textwidth]{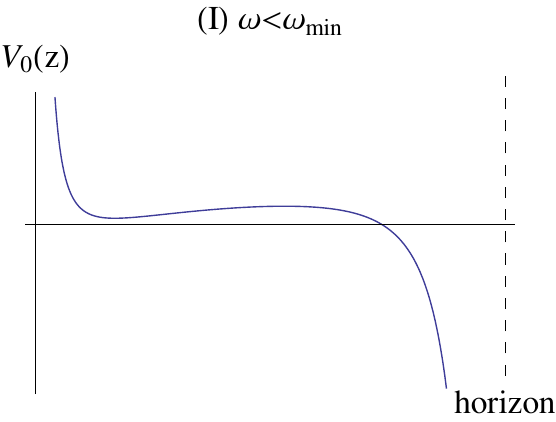}\quad
b) \includegraphics[width=0.35\textwidth]{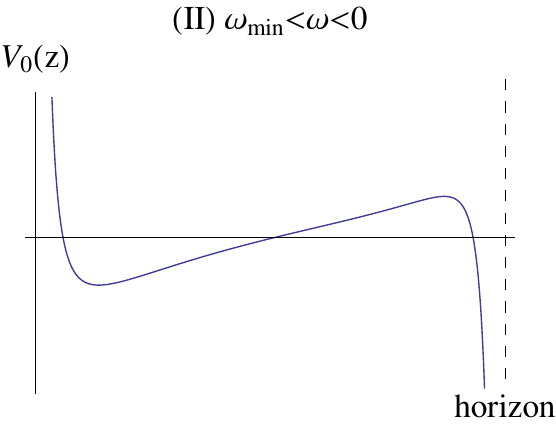} \\
c) \includegraphics[width=0.35\textwidth]{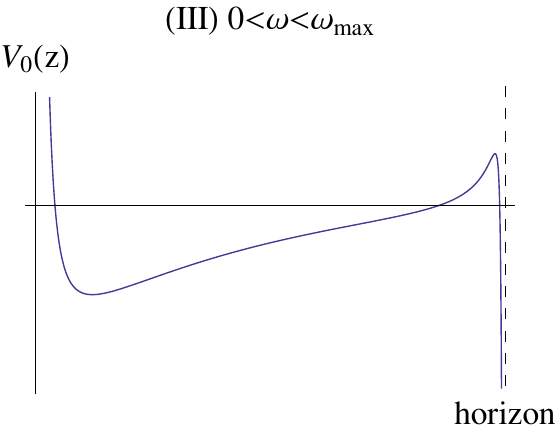}\quad
d) \includegraphics[width=0.35\textwidth]{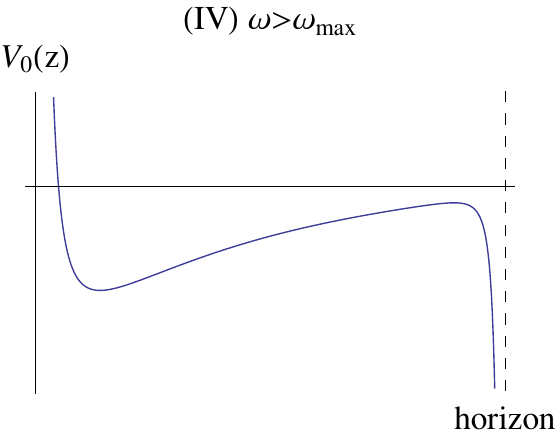}
\caption{\label{fig:Veff1} The leading order potential $V_0$ for the spinor or scalar when $\omega\neq 0$.
There exist quasi-bound states when $\omega_{\rm min} < \omega < \omega_{\rm max}$.
%There will be these three cases for both $k<k_F$ and $k>k_F$.
%When $m=0$, $V(z=0)$ will be finite, and the behavior of the wave
%function is determined by the boundary condition at $z=0$.
}
\end{figure}

\begin{figure}
%\end{minipage}\\[15pt]
%\begin{minipage}[t]{\textwidth}
\centering
a) \includegraphics[width=0.35\textwidth]{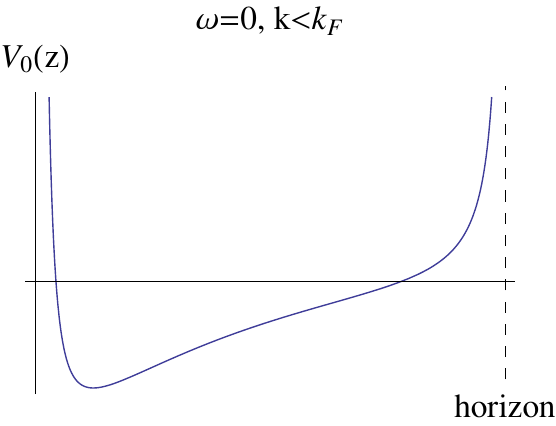}\qquad
b) \includegraphics[width=0.35\textwidth]{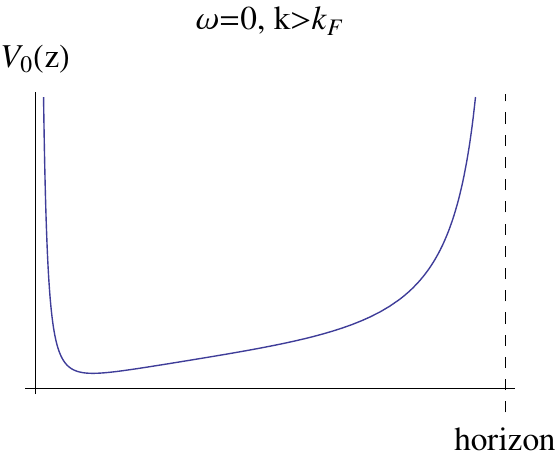}
\caption{\label{fig:Veff0} The leading order potential $V_0$ for the spinor or scalar
when $\omega=0$.
There exists a bound state provided $k$ is not too large.
%, but different from the extremal black hole.
}
%\end{minipage}
\end{figure}

The rough picture of the WKB analysis of the Schr\"odinger equation (\ref{Schreq}) with the spinor potential function $V_{k,m}$ (\ref{diracpot}) is easily explained.
Schematically, we may write our potential as
%At leading order in $\hbar$, the potential is
\begin{equation}
V_{k,m} = \frac{V_0}{\hbar^2} + \frac{V_1}{\hbar} + V_2 \,.
%V_0 = \frac{1}{\hbar^2} \left( g_{zz} m^2 - Z_k^2 Z_{-k}^2 \right) \ .
\end{equation}
Considering only the leading order term $V_0 = g_{zz} m^2 - Z_k Z_{-k}$, there is a barrier at the conformal boundary $z=0$ provided $m^2>0$.  At the horizon $z=1$, the potential $V_0$ is unbounded below provided $\omega^2 > 0$.  For intermediate values $0<z<1$ and appropriate choices of $k$, $\mu$, and $q$, one may find a potential well where $V_0 < 0$ separated from the horizon by a barrier where $V_0 > 0$ (see figure \ref{fig:Veff1}).  For discrete choices of $\omega$, the wave function will satisfy a Bohr-Sommerfeld type quantization condition, and
the potential well will support quasi-bound states.  Tunneling through the barrier to the horizon gives $\omega$ a small imaginary part.
Note in the special case $\omega=0$ shown in figure \ref{fig:Veff0}, the potential at the horizon becomes a barrier, suggesting that the quasinormal modes have very small imaginary part close to the origin.


The detailed picture of this WKB analysis is more intricate.  At leading order in $\hbar$, the Schr\"odinger potentials for the spinor and scalar are identical, and the results will be independent of the spin that is this paper's main focus.  Moreover, there is an important qualitative difference in the
QNM spectrum for the spinor and scalar that is not captured at this leading order.  The charged scalar QNMs may lie in the upper half of the complex $\omega$ plane signaling a perturbative instability \cite{Gubser:2008px, Hartnoll:2008vx},
while the spinor QNMs will not.
Yet, as we will see, the leading order WKB analysis would put the spinor QNMs in the upper half plane as well.

To capture the effects of spin, we will keep the first subleading term in the $\hbar$ expansion of $V_{k,m}$.
Our WKB wavefunction is then
\begin{equation}
\phi_{\rm WKB} = \frac{1}{V_0^{1/4}} \exp \left( \pm \frac{1}{\hbar} \int \sqrt{V_0} \left( 1 + \hbar \frac{V_1}{2 V_0} \right) dz \right).
\end{equation}
The classical turning points are defined in terms of the zeroes of $V_0$.

\begin{figure}
\center
\includegraphics[]{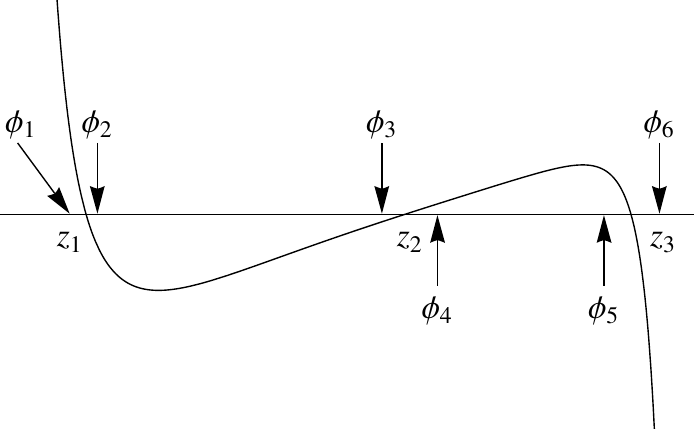}
\caption{\label{fig:wkb} A typical spinor or scalar potential at leading order in $\hbar$.  The classical turning points are $z_1$, $z_2$, and $z_3$.  The WKB wave functions $\phi_i$ correspond to
eqs.~(\ref{wkb1})--(\ref{wkb8}) }
\end{figure}

To set up the WKB problem, let $z_1 < z_2$ be the turning points bounding the classically allowed region.
Similarly, let $z_2 < z_3$ bound the potential barrier.  We begin by writing formal expressions for the WKB wave functions to the left and right of the three points $z_1$, $z_2$, and $z_3$ valid to next to leading order in $\hbar$:
\begin{eqnarray}
\phi_1 &=& \frac{1}{V_0^{1/4}} \left( A_1 e^{\frac{1}{\hbar} \int_z^{z_1}  \sqrt{V_0} \left( 1 + \hbar V_1 / 2 V_0 \right) dz} + B_1 e^{ -\frac{1}{\hbar}\int_z^{z_1}  \sqrt{V_0} \left( 1 + \hbar V_1 / 2 V_0 \right) dz} \right),
\label{wkb1}
\\
\phi_2 &=& \frac{1}{(-V_0)^{1/4}}\left( A_{2} e^{  \frac{i}{\hbar}\int_{z_1}^{z} \sqrt{-V_0} \left( 1 + \hbar V_1 / 2 V_0 \right)dz} +B_{2} e^{ - \frac{i}{\hbar} \int_{z_1}^{z} \sqrt{-V_0} \left( 1 + \hbar V_1 / 2 V_0 \right)dz} \right), \\
\phi_3 &=& \frac{1}{(-V_0)^{1/4}}\left( A_{3}e^{  \frac{i}{\hbar}\int_{z}^{z_2} \sqrt{-V_0}\left( 1 + \hbar V_1 / 2 V_0 \right) dz} + B_{3} e^{ - \frac{i}{\hbar}\int_{z}^{z_2} \sqrt{-V_0} \left( 1 + \hbar V_1 / 2 V_0 \right)dz} \right), \\
\phi_4 &=&  \frac{1}{V_0^{1/4}} \left(  A_4 e^{\frac{1}{\hbar} \int_{z_2}^{z} \sqrt{V_0} \left( 1 + \hbar V_1 / 2 V_0 \right)dz} + B_4 e^{ -\frac{1}{\hbar}\int_{z_2}^{z} \sqrt{V_0} \left( 1 + \hbar V_1 / 2 V_0 \right)dz} \right),
\\
\phi_5 &=&  \frac{1}{V_0^{1/4}} \left( A_5  e^{ \frac{1}{\hbar}\int_{z}^{z_3} \sqrt{V_0} \left( 1 + \hbar V_1 / 2 V_0 \right)dz} +B_5 e^{ -\frac{1}{\hbar}\int_{z}^{z_3} \sqrt{V_0} \left( 1 + \hbar V_1 / 2 V_0 \right)dz} \right),
\\
\phi_6 &=& \frac{1}{(-V_0)^{1/4}}\left( A_6 e^{ \frac{i}{\hbar}\int_{z_3}^{z} \sqrt{-V_0} \left( 1 + \hbar V_1 / 2 V_0 \right)dz} + B_6 e^{ - \frac{i}{\hbar}\int_{z_3}^{z} \sqrt{-V_0} \left( 1 + \hbar V_1 / 2 V_0 \right)dz} \right).
\label{wkb8}
\end{eqnarray}
Figure \ref{fig:wkb} portrays a typical potential with the regions labeled in which the six WKB wave functions are valid.

The wave functions in adjacent regions
will be related by connection matrices $M_i$ such that $v_{i+1} = M_i v_i$ where
$v_i = (A_i, B_i)^T$.
Given these connection matrices, we can obtain a semi-classical quantization condition on $\omega$ by applying boundary conditions.
We will take Dirichlet boundary conditions at $z=0$ that $A_1 = 0$.
At the horizon, we have ingoing boundary conditions that $\phi \sim (1-z)^{-i \omega / 3}$.
The WKB wave function (\ref{wkb8})
has the near horizon expansion $\phi_6  \sim A_6 (1-z)^{- i |\omega| / 3} + B_6 (1-z)^{i |\omega| / 3}$.
Thus when $\omega >0$ we should take $B_6 = 0$, and when $\omega < 0$ we should take instead $A_6 = 0$.
The equation
\begin{equation}
v_6 =  M_5 M_4 M_3 M_2 M_1 v_1
\label{quantformal}
\end{equation}
then provides a Bohr-Sommerfeld like quantization condition on $\omega$.

The matrices $M_i$ are all well known.
To go from a classically forbidden region to the classically allowed region, we use the standard WKB connection formula
\begin{equation}
M = \left(
\begin{array}{cc}
\frac{i}{2}  & 1 \\
\frac{1}{2}  & i
\end{array}
\right).
\end{equation}
Thus we find that $M_1 = M$ and $M_3 = M^{-1}$.
To go from $\phi_2$ to $\phi_3$ or from   $\phi_4$ to $\phi_5$, we make use of the fact that $\int_a^z f(z) dz = -\int_z^b f(z) dz + \int_a^b f(z) dz$:
\begin{equation}
M_2 =
\left(
\begin{array}{cc}
0 & e^{-i L} \\
e^{i L} & 0
\end{array}
\right),\qquad
M_4 =
\left(
\begin{array}{cc}
0 & e^{-K} \\
e^{K} & 0
\end{array}
\right),
\end{equation}
where
\begin{equation}
L \equiv \frac{1}{\hbar}\int_{z_1}^{z_2}\sqrt{-V_0} \left( 1 + \hbar \frac{V_1}{2 V_0} \right)dz \,, \qquad
K \equiv \frac{1}{\hbar}\int_{z_2}^{z_3}\sqrt{V_0} \left( 1 + \hbar \frac{V_1}{2 V_0} \right)dz  \,.
\end{equation}

The connection matrix $M_4$ deserves closer scrutiny because the integral $K$ is not always well defined.
From eq.\ (\ref{diracpot}), we see that the potential term $V_1$ (for the $\phi_+$ case) has the form
\begin{equation}
V_1 = - g_{zz} m \frac{ \partial_z (\sqrt{g^{zz}} Z_k)}{Z_k} \,.
\end{equation}
Thus $V_1$ will have a simple pole where $Z_k$ vanishes.
From the form of $V_0 = g_{zz} m^2 - Z_{-k} Z_k$, it is clear that $Z_k$ will only vanish in a classically forbidden region where $V_0 > 0$.  Thus, the integral $L$ will never be singular in this way.
Let $z=a_i$ be the locations of the simple poles of $V_1$.  Near $a_i$, the integrand for $K$ looks like
$\frac{1}{2} (z-a_i)^{-1}$.  We regulate the singularity by taking a small semi-circular detour in the complex $z$ plane.
The detour introduces a factor of $\pm i \pi /2$, depending on the choice of detour above or below the singularity.
This extra phase factor introduces a relative minus sign between the two nonzero entries of $M_4$.  We do not need to worry about the overall sign as it will not affect the quantization condition.

We can be more precise about where $Z_k$ has zeroes.
From (\ref{Zkeq}), $Z_k$ is manifestly positive in the region $0<z<1$ if both $\omega>0$ and $k<0$.  Thus in this case, there will be no singularity to worry about.  However, if $\omega<0$ there will in general be one such value $z=a_1$ and if $\omega > 0$ and $k>0$, there will be either zero or two such values.

Given the matrices $M_i$ and the relation (\ref{quantformal}),
we find the quantization condition on $\omega$ looks in general like
\begin{equation}
\cos L(\omega) + i c e^{-2 K(\omega)} \sin L(\omega) = 0 \,.
\label{quantcond}
\end{equation}
In deriving (\ref{quantcond}), we made the implicit assumption that the turning points lie on the real axis.  However, the quantization condition implies $\omega$ is not real, and if $\omega$ is not real,
the turning points will in general not lie on the real axis.
To get out of this apparent contradiction, we assume that $|c e^{-2  K}| \ll 1$.  Then the imaginary part of $\omega$ should be small, and we can use (\ref{quantcond}) to estimate it.  We find the QNMs at $\omega_n$, $n=1,2,3,\ldots$,  satisfy
\begin{equation}
L(\operatorname{Re} \omega_n ) = \pi (n - 1/2) \,, \qquad
\operatorname{Im} \omega_n \approx  \left. c \frac{e^{-2  K}}{dL/ d\omega} \right|_{\omega = \operatorname{Re}\omega_n}.
\end{equation}
In all our examples, it is also true that $dL/d\omega > 0$.  Thus the sign of the imaginary part is determined by the sign of
$c e^{-2K}$.

Specializing to the $\phi_+$ case for concreteness, there are several cases to consider.
\begin{itemize}

\item
When $\omega >0$ and $k<0$, the singularities at $z=a_i$ are absent, and we find for both the scalar and spinor the quantization condition (\ref{quantcond}) with  $c = -1/4$.
These QNMs lie in the lower half plane.
\item
For the scalar when $\omega<0$, the change in the WKB boundary conditions at the horizon
leads to the relation (\ref{quantcond}) with $c = 1/4$.
These QNMs lie in the upper half plane.

\item
For the spinor when $\omega < 0$, there is a pole in $V_1$ at $z=a_1$.  Deforming the contour to avoid the pole adds a phase factor $\pi i/2$ to the integral $K$.  Thus while $c=1/4$ as in the scalar case, $e^{-2K} <0$ is negative.
These QNMs lie in the lower half plane.

\item
We may also consider the case where $\omega > 0$ and $k>0$.  In this case, $Z_k$ may have no zeroes or two zeroes in the region $0<z<1$.  If there are no zeroes, we reduce to the $\omega>0$ and $k<0$ case.  If there are two zeroes, then $e^{-2K}>0$ and the QNMs are still in the lower half plane.

\end{itemize}

\begin{figure}
\centering
\includegraphics[]{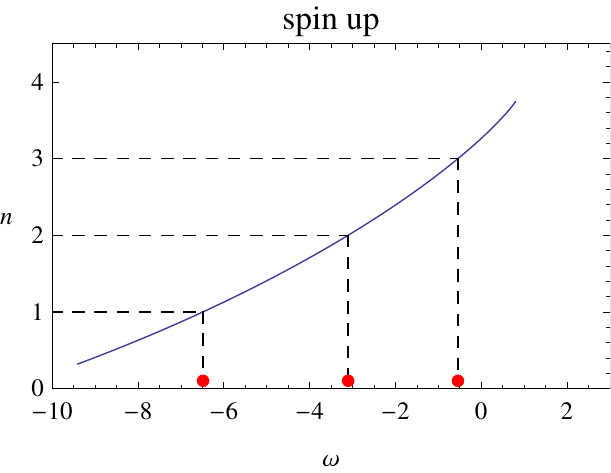}\qquad
\includegraphics[]{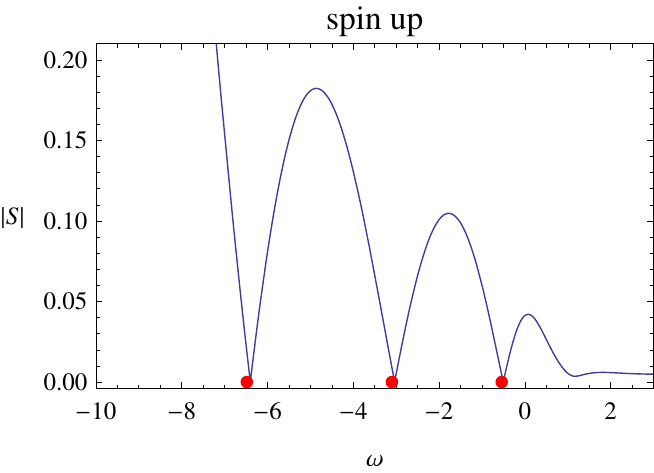}\\[8pt]
\includegraphics[]{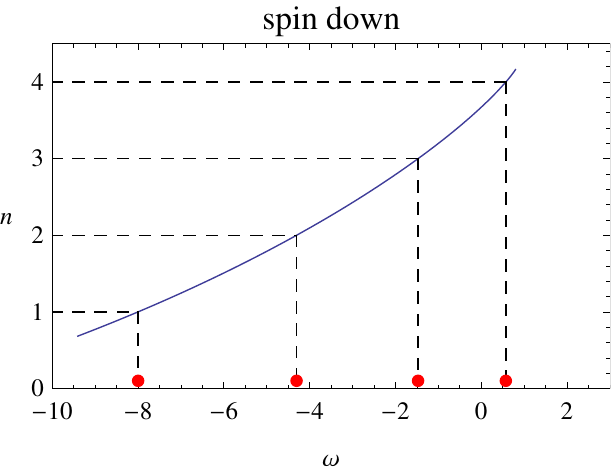}\qquad
\includegraphics[]{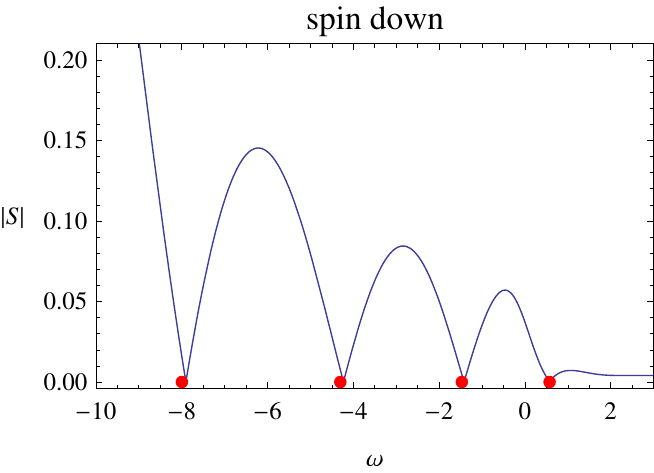}
\caption{\label{fig:nm} Normal modes obtained by the Bohr-Sommerfeld quantization, marked by red dots.
%The spin splitting has been captured by the WKB method we use.
%The WKB method without spin can only give the same modes for the two spins.
The parameters are $m=2$, $k=5$, and $\mu_q=25$. Here $|S|=|a_\alpha|$
is the source (denominator) of the Green's function.}
\end{figure}

At first order in $\hbar$, the WKB approximation already works pretty well.  Below are tables comparing the first order WKB against numerical integration of the Dirac equation.
For $m=2$ and $\mu_q = 25$ we have the following results when $k = \pm 5$, illustrating the spin splitting:
\begin{equation}
\label{mu25k5}
\begin{array}{|r|r|c|r|r|}
\cline{1-2} \cline{4-5}
\multicolumn{2}{|c|}{k = -5} &\hskip 0.3in & \multicolumn{2}{|c|}{k = 5}  \\
\cline{1-2} \cline{4-5}
\multicolumn{1}{|c|}{\rm numeric} & \multicolumn{1}{|c|}{\rm WKB} && \multicolumn{1}{|c|}{\rm numeric} & \multicolumn{1}{|c|}{\rm WKB}\\
\cline{1-2} \cline{4-5}
-7.90 - 0.00516  i &  -7.99 - 0.00504 i &&  -6.40 - 0.00398 i & -6.48 - 0.00383 i \\
-4.23 - 0.00597 i & -4.30 - 0.00587 i &&  -3.04 - 0.00476 i &  -3.10 - 0.00471 i\\
-1.42 - 0.00545 i &  -1.47 -0.0052 i &&  -0.504 - 0.00471 i &  -0.540 -0.0044 i\\
\cline{4-5}
0.574 - 0.0390 i  & 0.576 - 0.041 i\\
\cline{1-2} \end{array}
\end{equation}
We have presented the same information graphically in figure \ref{fig:nm}.

\begin{figure}
\centering
\includegraphics[width=0.382\textwidth]{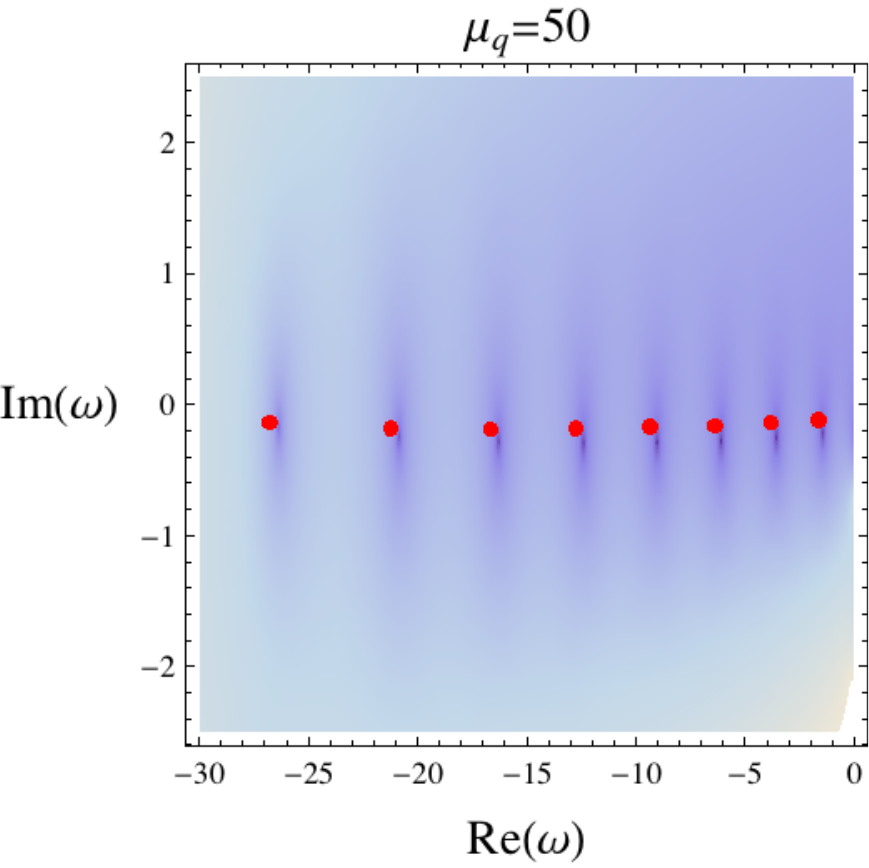}\qquad
\includegraphics[width=0.382\textwidth]{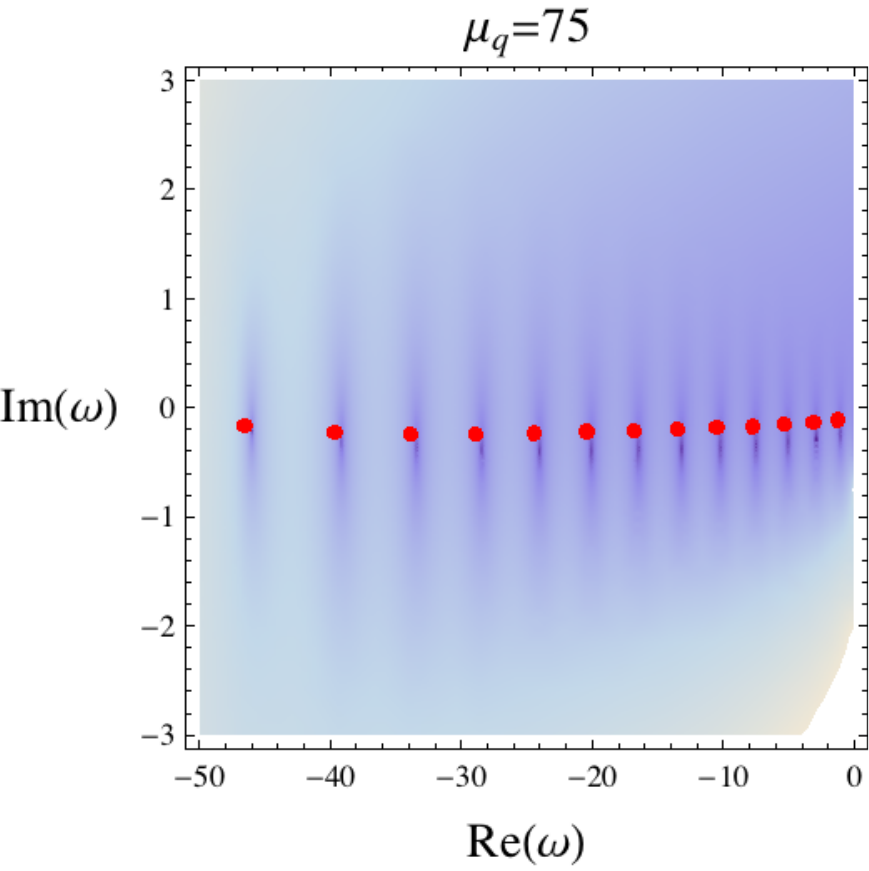}
\caption{\label{fig:QNM50} Quasibound states in the complex $\omega$ plane. The red dots are the quasinormal modes obtained by the generalized WKB formula. The parameters are $m=2$, $k=0$. Two values of $\mu_q$ are considered.
}
\end{figure}

For $m=2$, $k=0$, we also present results for $\mu_q = 50$ and $\mu_q = 75$.
\begin{equation}
\label{mu50mu75}
\begin{array}{|r|r|c|r|r|}
\cline{1-2} \cline{4-5}
\multicolumn{2}{|c|}{\mu_q = 50} &\hskip 0.3in & \multicolumn{2}{|c|}{\mu_q = 75}  \\
\cline{1-2} \cline{4-5}
\multicolumn{1}{|c|}{\rm numeric} & \multicolumn{1}{|c|}{\rm WKB} && \multicolumn{1}{|c|}{\rm numeric} & \multicolumn{1}{|c|}{\rm WKB}\\
\cline{1-2} \cline{4-5}
 -26.35 - 0.172 i & -26.77 - 0.135 i &&  -45.98 - 0.214 i & -46.50 - 0.165i\\
 -20.84 - 0.248 i & -21.25 - 0.18 i && -39.14 - 0.313 i & -39.64 - 0.227 i\\
 -16.29 - 0.281 i &  -16.66 -0.19 i  && -33.41 - 0.361 i &   -33.87 - 0.24 i\\
 -12.41 -0.292  i &  -12.74 -0.18 i && -28.43 - 0.385 i &  -28.87 - 0.24 i\\
-9.03 - 0.290 i &  -9.34 - 0.17 i  && -24.02 - 0.394 i &  -24.42 - 0.23 i\\
 -6.10 - 0.278 i & -6.37 - 0.16 i && -20.05 - 0.394 i  & -20.42 - 0.22 i\\
-3.57 - 0.255 i & -3.80 - 0.14 i && -16.45 - 0.388 i  & -16.80 - 0.21 i\\
 -1.44 - 0.218 i & -1.63 - 0.12 i && -13.18 - 0.377 i&  -13.51 - 0.195 i\\
 \cline{1-2}
\multicolumn{3}{c|}{} & -10.20 - 0.362 i & -10.50 - 0.18 i\\
 \multicolumn{3}{c|}{} & -7.50 - 0.343i & -7.77 - 0.17 i\\
 \multicolumn{3}{c|}{} &  -5.06 - 0.319 i & -5.31 - 0.15 i\\
\multicolumn{3}{c|}{} & -2.90 - 0.288 i &  -3.11 - 0.13 i \\
\multicolumn{3}{c|}{} &  -1.05 - 0.240 i &  -1.23 - 0.11 i \\
\cline{4-5}
\end{array}
\end{equation}
The results are presented graphically in figure \ref{fig:QNM50}.

We will stop at first order in the WKB approximation, but there are some interesting complications that appear at second order.
 There exist second order poles in $V_{k,m}$ at $z=0$, 1 and $a_i$ that require a more careful consideration.
Regardless of the background, the second order poles near $z=a_i$ have the universal form
\begin{equation}
V_{k,m} \approx \frac{3/4}{(z-a_i)^2} \,.
\end{equation}
In our particular case, at the conformal boundary, we find that
\begin{equation}
V_{k,m} = \frac{m^2 - m \hbar }{\hbar^2 z^2} + O(z^{-1})  \,.
\end{equation}
At the horizon, we have instead
\begin{equation}
V_{k,m} = - \left( \frac{1}{4} + \frac{16 \omega^2 }{\hbar^2(12 - \mu^2)^2} \right) \frac{1}{(1-z)^2} + O(1-z)^{-3/2} \,.
\end{equation}
The issue with second order poles in the potential  is that the naive WKB wave functions do not have the proper scaling behavior near such points.  Langer \cite{Langer} proposed a modification, justified by a rescaling of the wave function and redefinition of $z$, that boils down to adding by hand a term of the form $\frac{1}{4} (z-z_s)^{-2}$ to the potential for each second order pole $z_s$.  These Langer modification factors are second order in our $\hbar$ expansion, and thus we have neglected them.

\section{Discussion}

We have tried to show in this paper that spin affects in important ways the physics of holographic constructions involving fermions.  The pole of the field theory fermionic Green's function obeys a Rashba type dispersion relation, indicating the importance of spin orbit coupling in the gravitational side of the construction.  The spin also plays an important role in determining the lifetimes of quasiparticles.  Without the spin corrections in the WKB approximation, the fermionic dispersion relation would have an imaginary part of the wrong sign!
That spin plays an important role in the bulk is ironic given that in a 2+1 dimensional field theory dual, we argued in the introduction that the fermions should be both massless and (hence) spinless.

This work leaves open several issues that we would like to return to at some point.  One is a more thorough exploration of the parameter space of our model.  We would like to understand better the qualitative difference between bulk fermions with large masses $mL \gtrsim 1$ and small masses $mL \lesssim1$ illustrated by figures \ref{fig:m2p}a and \ref{fig:m0}.  While the gravity explanation is related to the absence of a rest frame for a relativistic particle, the field theory interpretation is less clear.   The limit $\omega \to 0$ played an important role in previous works on the subject (see for example \cite{fau09}), and we would like to see what our WKB formalism predicts for the magnitude of the imaginary part of the dispersion relation and the corresponding lifetime of the quasiparticles.   (How does including $\hbar$ corrections change the results of \cite{har11b}?)  Third, it would be interesting to consider the $T \to 0$ limit in more detail.

It would also be interesting to work with a wider variety of backgrounds, for example the electron star or the holographic superconductor.  We believe the qualitative nature of our story involving spin-orbit coupling and quasinormal mode placement will not change, but there may be other interesting effects.  For example, with the d-wave superconductor studied in \cite{Benini:2010pr}, it was precisely this spin-orbit coupling which gave rise to Fermi arcs.

%[[ things to think about:
%
%\begin{itemize}
%
%\item
%standard validity of WKB checks
%
%\end{itemize}
%
%]]

\section*{Acknowledgements}
We thank  K.~Balasubramanian, D.~Hofman, T.~Faulkner, S.~Hartnoll, R.~Loganayagam, L.~Rastelli,
and D.~Vegh for discussions.
This work was supported in part by the National Science Foundation under Grants No. PHY-0844827 and PHY-0756966.
CH thanks the Sloan Foundation for partial support, and the KITP for hospitality (and partial support under NSF Grant No.\ PHY-1125915).

\appendix

\section{Analytic results from Heun polynomials}
\label{app:Heun}

In the AdS/CFT correspondence, some perturbation equations can be written in terms of the Heun equation, which has four regular singularities. Under certain conditions, the solution of the Heun equation is a polynomial, which can help us to obtain an exact solution to the Green's function. The Heun differential equation is
\begin{equation}
\frac{d^2y}{dx^2}+\left(\frac{\gamma}{x}+\frac{\delta}{x-1}+\frac{\epsilon}{x-a}\right)\frac{dy}{dx}+\frac{\alpha\beta x-Q}{x(x-1)(x-a)}y=0\,,
\end{equation}
where $\alpha+\beta+1=\gamma+\delta+\epsilon$ \cite{ron95}. The regular singularities are at $x=0$, $1$, $a$, and $\infty$. The solution to this equation is called the Heun function, if it is regular at both $x=0$ and $x=1$ (assuming $a>1$). We denote the Heun function by ${\rm HeunG}(a,Q;\alpha,\beta,\gamma,\delta;x)$, which is symmetric in $\alpha$ and $\beta$.
The Heun function can be written as a power series ${\rm HeunG}(x) = \sum_{r=0}^\infty c_r x^r$ where the coefficients satisfy a three term recursion relation \cite{ron95}:
\[
(r-1+\alpha)(r-1+\beta)c_{r-1}-\{r[(r-1+\gamma)(1+a)+a\delta+\epsilon]+Q\}c_r+(r+1)(r+\gamma)ac_{r+1}=0 \,.
\]
Given this recursion relation, it is clear that if $\alpha$ or $\beta = -n$ and if $c_{n+1} = 0$, then ${\rm HeunG}(x)$ is an $n$th order polynomial.\footnote{%
 There are other types of ``Heun polynomials" \cite{ron95}.
 For example, another solution to the Heun equation is
 $x^{1-\gamma}{\rm HeunG}(a,(a\delta+\epsilon)(1-\gamma)+Q;\alpha+1-\gamma,\beta+1-\gamma,2-\gamma,\delta;x)$,
 which indicates that the HeunG on the right-hand side is a polynomial of order $n$
 when $\alpha+1-\gamma=-n$. In this work, other polynomials give unphysical
 results or the same results as the case $\alpha=-n$.
}
In general, the condition $c_{n+1}=0$ is an $(n+1)$th order algebraic equation in the parameters of the Heun function.
Since we need to solve an $(n+1)$th-order algebraic equation to obtain an $n$th-order Heun polynomial, we usually cannot obtain explicit solutions when $n\geq 4$. (For comparison, the series expansion for the hypergeometric function $_2F_1(\alpha,\beta,\gamma;x)$ is defined by a two term recursion relation, and the polynomial condition is
just $\alpha=-n$.) The set of solutions $c_{n+1}=0$ may include unphysical regions of parameter space, for example regions where the charge or momentum is imaginary.  We will need to further restrict the solution set.

Setting
\begin{equation}
\psi = \left(
\begin{array}{c}
u_1 \\
u_2
\end{array}
\right),
\end{equation}
the two coupled equations (\ref{eq:dirac2}) for $u_1$ and $u_2$
%\eqref{eq:u1} and \eqref{eq:u2}
can be reduced to a single second-order ODE. To avoid the awkward square root, we define $u_\pm=u_1\pm u_2$, and then the Dirac equations are
\begin{eqnarray}
\partial_zu_+-\frac{i(\omega+qA_t)}{f}u_+ &=& \frac{m+ikz}{z\sqrt{f}}u_-\,,\\
\partial_zu_-+\frac{i(\omega+qA_t)}{f}u_- &=& \frac{m-ikz}{z\sqrt{f}}u_+\,.\label{eq:upm}
\end{eqnarray}
The equation for $u_-$ is given by\footnote{%
 Note that this equation can be put in Schr\"odinger form and gives an alternate starting point for a WKB approximation.  The corresponding Schr\"odinger potential is complex which makes phase integral methods substantially more involved.
}
\begin{eqnarray}
u_-''+\left(\frac{f'}{2f}+\frac{m}{z(m-ikz)}\right)u_-'
+\left[\frac{(\omega+qA_t)^2}{f^2}-\frac{m^2}{z^2f}-\frac{k^2}{f}\right.\nonumber\\
\left.+\frac{i(\omega+qA_t)}{f}\left(-\frac{f'}{2f}+\frac{m}{z(m-ikz)}\right)+\frac{iqA_t'}{f}\right]u_-=0\,.
\end{eqnarray}
We need to solve this equation with in-falling wave boundary condition at the horizon, and then plug $u_-$ into eq.~\eqref{eq:upm} to obtain $u_+$.

For the massless fermion in $AdS_5$, we obtain a second-order ODE with four regular singularities as follows:\footnote{%
 While we have focused on $AdS_4$ backgrounds in the bulk of the paper, with an appropriate choice of gamma matrices \cite{gul10}, eq.~(\ref{eq:dirac2}) also holds for $AdS_5$.
}
\begin{equation}
u_-''+\frac{f'}{2f}u_-'+\left[\frac{(\omega+qA_t)^2}{f^2}-\frac{k^2}{f}-\frac{if'(\omega+qA_t)}{2f^2}
+\frac{iqA_t'}{f}\right]u_-=0\,,
\end{equation}
where $f=1-z^4$ and $A_t=1-z^2$. The solution is
\begin{equation}
\begin{split}
u_- &\sim (z+1)^{i\omega/4+1/2}(z-1)^{-i\omega/4+1/2}(z+i)^{-q/2+1/2-\omega/4}(z-i)^{-3/2+q/2+\omega/4}\\
& \times{\rm HeunG}\left(\frac{1}{2},\frac{3-(1-i)\omega}{2}-\frac{ik^2}{4}-q;2,\frac{3-\omega}{2}-q,
\frac{3+i\omega}{2},\frac{3-i\omega}{2};\frac{(1-i)(z+1)}{2(z-i)}\right).
\end{split}
\end{equation}
We can see that when
\begin{equation}
\omega=2n+3-2q\,,\qquad n=0,1,2,\cdots,
\end{equation}
we can obtain an $n$th-order polynomial solution $p_n$,
\begin{equation}
{\rm HeunG}(x)=p_n=\sum_{r=0}^nc_rx^r\,,\qquad x=\frac{(1-i)(z+1)}{2(z-i)}\,,
\end{equation}
if $k$ satisfies an $(n+1)$th-order algebraic equation.

The zeroth-order polynomial is
\begin{equation}
p_0=1\,,\qquad {\rm when}\quad k^2=6-4q\,.
\end{equation}
With this exact solution, we can obtain the Green's function
\begin{equation}
G(\omega=3-2q,k=\sqrt{6-4q},q)=\frac{4-2q-\sqrt{6-4q}}{4-2q+\sqrt{6-4q}}i\,.
\end{equation}
For the other sign of $k$, $G(-k)=-1/G(k)$. We can see that ${\rm Im}(G)>0$ is always satisfied (if $k$ is real). The first-order polynomial is
\begin{equation}
p_1=1-\frac{2\omega-2i\omega-6+ik^2+4q}{i\omega+3}x\,,\qquad {\rm when}\quad k^2=15-6q\pm\sqrt{4q^2-20q+33}\,.
\end{equation}
The Green's function
\begin{equation}
G(\omega=5-2q,k=\sqrt{15-6q\pm\sqrt{4q^2-20q+33}},q)
\end{equation}
can be exactly expressed. If we take the plus sign, for example, the result is
\begin{small}
\begin{equation}
\frac{15-4q+(7-2q)\sqrt{4q^2-20q+33}-(3+\sqrt{4q^2-20q+33})\sqrt{15-6q+\sqrt{4q^2-20q+33}}}
{15-4q+(7-2q)\sqrt{4q^2-20q+33}+(3+\sqrt{4q^2-20q+33})\sqrt{15-6q+\sqrt{4q^2-20q+33}}}i\,.
\end{equation}
\end{small}
To obtain the second-order polynomial, we need to solve a third-order algebraic equation for $k$. There are higher-order polynomials, and the results are more complicated. The denominator of the above two Green's functions are non-zero for all real $q$. It is unlikely that we can solve for a Fermi momentum $k_F$ in this way.

The exact results can be used to check the numerical program. For example, some exact results are
\begin{eqnarray}
G(\omega=1,k=\sqrt{2},q=1) &=& (3-2\sqrt{2})i\,,\\
G(\omega=1,k=\sqrt{6},q=2) &=& \frac{59-24\sqrt{6}}{5}i\,,\\
G(\omega=-1,k=0,q=3) &=& i\,,
\end{eqnarray}
where the last one is consistent with $G(\omega,k=0)=i$ for massless spinors \cite{fau09}. As more examples, we list the numerical value of the exact result and the numerical result in some special cases as follows:
\begin{center}
\begin{tabular}{|c|c|c|c|c|}
\hline
$\omega$ & $k$ & $q$ & exact & numerical\\\hline
$7$ & $\sqrt{21+\sqrt{57}}$ & $-1$ & $0.21336115i$ & $0.21336116i$\\\hline
$5$ & $\sqrt{15+\sqrt{33}}$ & $0$ & $0.16186619i$ & $0.16186620i$\\\hline
$3$ & $\sqrt{9+\sqrt{17}}$ & $1$ & $0.10121094i$ & $0.10121094i$\\\hline
$1$ & $\sqrt{6}$ & $2$ & $0.04244923i$ & $0.04244924i$\\\hline
$0$ & $2^{3/4}$ & $5/2$ & $0.04177353i$ & $0.04177353i$\\\hline
\end{tabular}
\end{center}

There is a hidden algebraic structure in the Heun equation. A representation of the SU(2) algebra is
\begin{eqnarray}
S_+ &=& z^2\partial_z-2sz\,,\\
 S_- &=& -\partial_z \,,  \\
%S_- &=& \partial_+,\\
S_0 &=& z\partial_z-s\,,
\end{eqnarray}
where $[S_+,S_-]=2S_z$, $[S_0,S_\pm]=\pm S_\pm$, and ${\bf S}^2=s(s+1)$.
If we consider the following Hamiltonian problem $H y(z) = E y(z)$ with
%If we substitute the above representation to the universal enveloping algebra of ${\rm SU}(2)$ up to the quadratic order:
\begin{equation}
H=\mathop{\sum_{i,j=-,0,+}}_{i\geq j}  a_{ij}S_iS_j+\sum_{i=+,0,-}b_iS_i\,,\label{eq:H}
\end{equation}
we obtain a differential equation with polynomial coefficients \cite{tur88}:
\begin{equation}
P_4\,y''(z)+P_3\,y'(z)+P_2\,y(z)= E y(z)\,,
\end{equation}
where
\begin{align}
P_4 &=a_{++}z^4+a_{+0}z^3+(a_{00}-a_{+-})z^2-a_{0-}z+a_{--}\,,\\
P_3 &=2(1-2s)a_{++}z^3+[(1-3s)a_{+0}+b_+]z^2+[2s(a_{+-}-a_{00})+a_{00}+b_0]z \\
& \nonumber
\hspace{10mm} +sa_{0-}-b_-\,,\\
P_2 &=2s(2s-1)a_{++}z^2+2s(sa_{+0}-b_+)z+s^2a_{00}-s b_0\,.
\end{align}
There is a correspondence between the Heun equation and the spin system by the following identification:
\begin{align}
& a_{++}=0\,,\qquad a_{+0}=1\,,\qquad a_{--}=0\,,\\
& a_{00}-a_{+-}=-(1+a)\,,\qquad a_{0-}=-a\,,\\
%& 2s(a_{+-}-a_{00})+a_{00}+b_0=-(\alpha+\beta+1+t\gamma+(t-\gamma)\delta),\\
&  2s(a_{+-}-a_{00})+a_{00}+b_0 = - ( \gamma(1+a) + \delta a + \epsilon ) \,,  \\
& sa_{0-}-b_-=a\gamma\,,\qquad s^2a_{00} - s b_0 - E = Q\,,\\
& b_+-3s=\alpha+\beta\,,\qquad 2s(s-b_+)=\alpha\beta\,. \label{eq:spin}
\end{align}
Solving eq.~\eqref{eq:spin} gives $s=-\alpha/2$ or $s=-\beta/2$. If the total spin $s$ is an integer or half-integer, the $H$ in eq.~\eqref{eq:H} has a finite dimensional representation. Therefore, the eigenvalues satisfy an algebraic equation by diagonalizing $H$. This algebraic equation is equivalent to the condition for the existence of a Heun polynomial solution. If $s$ is not an integer or half-integer, the Hilbert space is infinite dimensional.

The interpretation of why there exist some exact solutions to the Green's function is as follows. When $m=0$, in the three-dimensional parameter space ($\omega$, $k$, $q$), there are discrete one-dimensional lines (subspaces) that are labeled by $n=1$, $2$, $\cdots$. On these lines, the Hilbert space (accidentally) becomes finite dimensional. So the eigenvalues satisfy an $n$th-order algebraic equation. When $n\leq 4$, we can explicitly solve the algebraic equation.

\end{document}